\documentclass[preprint,12pt,sort&compress,times]{elsarticle}
\journal{Physics Letters B}
\usepackage[breaklinks,colorlinks=true]{hyperref}
\usepackage{amsmath}
\usepackage{amssymb}
\usepackage{bm}
\usepackage{physics}

\newcommand{\rme}{\mathrm{e}}
\newcommand{\rmd}{\mathrm{d}}
\newcommand{\rmi}{\mathrm{i}}
\newcommand{\Nc}{N_{\text{c}}}
\newcommand{\Nf}{N_{\text{f}}}
\newcommand{\Tc}{T_{\text{c}}}
\newcommand{\iomega}{\Omega_{\text{I}}}
\newcommand{\biomega}{{\bm\Omega}_{\text{I}}}
\newcommand{\tiomega}{\tilde{\Omega}_{\text{I}}}

\newcommand{\Det}{\mathrm{Det}}

\newcommand{\kp}{k_\perp}
\newcommand{\bphi}{{\bm\phi}}
\newcommand{\balpha}{{\bm\alpha}}

\newcommand{\hamma}{\hat{\gamma}}

\begin{document}
\begin{frontmatter}

\title{Inhomogeneous confinement and chiral symmetry breaking induced by imaginary angular velocity}

\author[1]{Shi Chen}
\author[2]{Kenji Fukushima}
\author[2]{Yusuke Shimada}

\affiliation[1]{organization={School of Physics and Astronomy, University of Minnesota},
 addressline={Minneapolis},
 city={MN},
 postcode={55455},
 country={USA}}

\affiliation[2]{organization={Department of Physics, The University of Tokyo},
 addressline={7-3-1 Hongo, Bunkyo-ku},
 city={Tokyo},
 postcode={113-0033},
 country={Japan}}
 
\begin{abstract}
 We investigate detailed properties of imaginary rotating matter with gluons and quarks at high temperature.  Previously, we showed that imaginary rotation induces perturbative confinement of gluons at the rotation center.  We perturbatively calculate the Polyakov loop potential and find inhomogeneous confinement above a certain threshold of imaginary angular velocity.  We also evaluate the quark contribution to the Polyakov loop potential and confirm that spontaneous chiral symmetry breaking occurs in the perturbatively confined phase.
\end{abstract}

\end{frontmatter}

%%%%%%%%%%
\section{Introduction}

The physical mechanism and the interplay of confinement and chiral symmetry breaking are long-standing puzzles in Quantum Chromodynamics (QCD) that is a fundamental theory in terms of quarks and gluons.  At finite temperature, if the quark mass is infinitely large to quench dynamical quarks, confinement and deconfinement of gluons are classified by center symmetry.  Dynamical quarks explicitly break center symmetry~\cite{Svetitsky:1985ye}, and the order parameter for center symmetry, i.e., the Polyakov loop is only an approximate measure of confinement.  In the limit of zero quark mass, chiral symmetry is exact and the chiral condensate characterizes the state of matter.  Various phases of QCD have been considered as functions of external environmental parameters such as the temperature $T$, the quark chemical potential $\mu$, the isospin chemical potential~\cite{Son:2000by,Son:2000xc}, the electric and magnetic fields~\cite{Yamamoto:2012bd,Andersen:2014xxa,Chen:2017xrj}, and the rotation angular velocity $\omega$~\cite{Jiang:2016wvv,Wang:2018sur,Chernodub:2016kxh,Chen:2023cjt,Sun:2023kuu, Fujimoto:2021xix, Chen:2020ath,Braga:2022yfe,Yadav:2022qcl, Wang:2024szr, Yamamoto:2013zwa,Braguta:2020biu,Braguta:2021jgn,Yang:2023vsw, Mameda:2023sst, Sun:2024anu, Cao:2023olg, Jiang:2023hdr, Gaspar:2023nqk, TabatabaeeMehr:2023tpt, Chernodub:2020qah,Chernodub:2022veq,Braguta:2023iyx}, and their mixture~\cite{Chen:2015hfc,Flachi:2017vlp,Zhang:2018ome}.  In this way, the QCD phase diagram has been intensively studied in theoretical and experimental contexts~\cite{Fukushima:2010bq}.

%size of the system~\cite{Ebihara:2016fwa,Zhang:2020hha,Chernodub:2017ref}
% KF: finite size cannot control the phase diagram

Among various parameters, the angular velocity $\omega$ has attracted much attention in recent years.  An extraordinary value of $\omega \sim 10^{22}\ \mathrm{s}^{-1}$ is estimated based on the data in the heavy-ion collision~\cite{STAR:2017ckg}.  In nonrelativistic theories rotation effects are similar to those induced by magnetic fields, while relativistic rotation of quark matter has features analogous to finite density~\cite{Chen:2015hfc,Wang:2018sur,Chernodub:2016kxh,Fukushima:2020ncb}.  In the same way as chiral symmetry restoration at high density, chiral effective models such as the (Polyakov--)Nambu--Jona-Lasino model exhibit the chiral phase transition induced by rotation~\cite{Jiang:2016wvv,Wang:2018sur,Chernodub:2016kxh,Chen:2023cjt,Sun:2023kuu}.  Interestingly, rotation directly affects gluons, which makes a contrast to effects of finite density and magnetic fields, and the angular velocity turns out to be a useful probe to confinement and deconfinement.  For the purpose to investigate a chance of (de)confinement caused by rotation, the hadron resonance gas model and the holographic QCD model have been adopted~\cite{Sun:2023kuu,Fujimoto:2021xix,Chen:2020ath,Braga:2022yfe,Yadav:2022qcl, Wang:2024szr}, which predicted that real rotation favors deconfinement.  Thus, it is expected that the deconfinement critical temperature should decrease for larger $\omega$.

The numerical results from lattice-QCD simulations~\cite{Braguta:2020biu,Braguta:2021jgn,Yang:2023vsw} have invoked controversies reporting a trend opposite to the model predictions.  That is, the Polyakov loop decreases as a result of real rotation, so that the deconfinement critical temperature should increase.  A technical subtlety lies in the treatment of analytical continuation;  the sign problem can be evaded with \textit{imaginary} angular velocity, $\iomega$, and then the critical temperature is inferred from $\Tc(\omega^2)=\Tc(-\iomega^2)$ if the Wick rotation does not hit singularities.

The subtle connection between real and imaginary rotation has been studied recently~\cite{Chen:2023cjt,Yang:2023vsw,Cao:2023olg,Jiang:2023hdr,Chernodub:2022veq,Braguta:2023iyx} but the validity condition for analytical continuation is only partially clarified for the moment.  There are various scenarios; some support the lattice results~\cite{Mameda:2023sst, Sun:2024anu}, some propose a modified model which can explain the lattice results~\cite{Cao:2023olg}, some suggest a non-monotonic function of $\Tc(\omega)$~\cite{Jiang:2023hdr, Gaspar:2023nqk}.

Previously in Ref.~\cite{Chen:2022smf}, we exploited the perturbative Polyakov loop potential to investigate rotation effects based on the first-principles approach.  Thanks to the asymptotic freedom, QCD or the pure gluonic theory can be perturbatively analyzed if the temperature is sufficiently high.  Without rotation, the perturbative Polyakov loop potential, known as the GPY-Weiss potential~\cite{Gross:1980br,Weiss:1980rj,Weiss:1981ev,KorthalsAltes:1993ca,Gocksch:1993iy}, spontaneously breaks center symmetry, which is consistent with deconfinement expected at high temperature; for a review, see Ref.~\cite{Fukushima:2017csk}.  We obtained the Polyakov loop potential in the rotating pure gluonic system and found that the Wick rotation, $\iomega = -\rmi\omega$, is hindered by singularity which corresponds to the causality violation.  To cure this obstacle, the system must have a finite boundary, and then a closed analytical expression is no longer available.

Interestingly enough, the theory with imaginary angular velocity contains rich physics.  In our previous work~\cite{Chen:2022smf}, we discovered exotic realization of confinement even at arbitrarily high temperature, i.e., the perturbatively confined phase.  Thus, $\iomega$ is a novel theoretical device to tackle the confinement mechanism.  In the present work, we will discuss two nontrivial extensions from our previous study.  One is exploring inhomogeneous structures away from the imaginary rotation center.  In Ref.~\cite{Chen:2022smf}, we focused on $r=0$ only (where $r$ is the radial distance) and argued that confinement is homogeneously realized for $SU(2)$ at $\iomega/T=\pi$.  There are some related works~\cite{Chernodub:2020qah,Chernodub:2022veq,Braguta:2023iyx} in favor of inhomogeneous confinement/deconfinement with real and imaginary rotation.  We will show that the perturbative Polyakov loop potential exhibits an inhomogeneous distribution of the Polyakov loop at $r\neq 0$.  Another extension is the inclusion of dynamical quarks, with which we can consider chiral symmetry breaking in the perturbatively confined phase.  The relation between confinement and chiral symmetry breaking is not fully understood, but under reasonable assumptions, one can see that confinement leads to chiral symmetry breaking or spontaneous generation of quark mass.  We will verify that quark mass is indeed nonzero in the perturbatively confined phase.

%%%%%%%%%%
\section{Inhomogeneous confinement and deconfinement}

Real rotation is a real-time effect, while imaginary rotation is interpreted as a geometrical effect.  The latter has a theoretical advantage as explicated below.  In the presence of an angular velocity vector, $\bm{\omega}$, a thermal system is described by the following partition function:
\begin{equation}\label{eq:partition}
    \mathcal{Z}_{T,\bm{\omega}} = \tr\rme^{-\beta(\hat{H} - \bm{\omega}\cdot\hat{\bm{J}})}\,,
\end{equation}
where $\beta=1/T$ is the inverse temperature and $\hat{\bm{J}}$ denotes the total angular momentum operator.  We can regard the above expression as a thermal expectation value of the topological operator, $\rme^{\beta\bm{\omega}\cdot\hat{\bm{J}}}$.  For pure imaginary $\bm{\omega} = \rmi \biomega$, this topological operator,
\begin{equation}
    \rme^{ \rmi \beta \biomega\cdot\hat{\bm{J}}}\,,
\end{equation}
is a unitary operator that generates a rotation by $\beta|\biomega|$ angle along the rotation axis $\parallel\!\biomega$.  In the Euclidean path integral, we can take account of $\biomega$ in the partition function~\eqref{eq:partition} by imposing an aperiodic thermal boundary condition which in cylindrical coordinates takes the form,
\begin{equation}\label{eq:BC}
    % (\bm{x},\,\tau) \sim (\rme^{-\beta\biomega\times}\bm{x},\,\tau+\beta)
    (r,\theta,z,\tau) \sim (r,\theta+\beta\iomega,z,\tau+\beta)\,.
\end{equation}
% Here, $\biomega\times$ is regarded as a $\mathfrak{so}(3)$ generator.  
Alternatively, we can perform the coordinate transformation to change the condition~\eqref{eq:BC} to the ordinary periodic one, but then the metric gains nontrivial components in a rotating frame, i.e.,
\begin{align}
    g_{\mu\nu} =
    \begin{pmatrix}
    1 & 0 & 0 & 0\\
    0 & r^2 & 0 & r^2 \iomega\\
    0 & 0 & 1 & 0 \\
    0 & r^2\iomega & 0 & 1 + r^2 \iomega^2
    \end{pmatrix}\,.
\end{align}
% In both approaches the equations of motion take the identical form.
Both approaches lead us to the same partition function.

We shall discuss inhomogeneous structures in the perturbatively confined phase in $SU(\Nc)$ pure gluonic matter with $\Nc=2$ and $3$.  Although we have already given a derivation in Ref.~\cite{Chen:2022smf}, we quickly review the formulation to make this paper self-contained.  In our calculation, we take the cylindrical coordinates, $(r, \theta, z, \tau)$, and assume rigidly rotating matter along the $z$-axis.

We take a constant $A_{\text{B}4}$ background and then $\partial_\tau$ is replaced by the covariant derivative $D_\tau$.  The inhomogeneity will be studied within the local density approximation (in which the spatial derivatives on $A_{\text{B}4}$ are neglected).  Using the Cartan subalgebra $\mathfrak{h}$ of Lie algebra $\mathfrak{g}$ of gauge group $G$, the covariant derivative is
\begin{equation}
    D_\tau = \partial_\tau + \rmi\frac{\bm{\phi}\cdot\bm{H}}{\beta}\, ,
\end{equation}
where $\bm{H}$ is a vector of bases of $\mathfrak{h}$ and $\bphi\cdot\bm{H}$ is normalized $A_{\text{B}4}$.  Using this background field, we fix the gauge by setting $D_\mu A_\mu = 0$.

In the actual calculations, the scalar Laplacian, $- D^2_{\mathrm{s}} = - \qty(D_\tau -  \iomega\partial_\theta)^2 - r^{-1}\partial_ r \qty( r \partial_r ) - r^{-2}\partial_\theta^2 - \partial_z^2$, is the basic building block.  We solve the equation of motion, $-D^2_{\mathrm{s}} \Psi = \lambda \Psi$, to find the spectrum as
\begin{align}
    \lambda_{n,l,k,\bm{\alpha}}=\qty(\frac{2\pi {n} + \bm{\phi}\!\cdot\!{\bm{\alpha}}}{\beta} - \iomega{l} )^2 + {\kp}^2 + {k_z}^2 \,,
%    \qty(n,m\in\mathbb{Z}\,,\ \kp\in\mathbb{R}^+\,,\ k_z\in\mathbb{R}\,,\bm{\alpha}\in\Phi)\,.\no
\end{align}
where quantum numbers are $n,l\in\mathbb{Z}$, $\kp\in\mathbb{R}^+$, $k_z\in\mathbb{R}$, and $\bm{\alpha}\in\Phi$ with $\Phi$ denoting the union of the $\mathfrak{su}(\Nc)$ root system and the zero roots.
The ghost contribution needs the determinant of $-D^2_{\mathrm{s}}$, which is given in terms of $\lambda_{n,l,k,\bm{\alpha}}$.
%\begin{subequations}
 %   \begin{align}
  %      \Psi_{n,m,k,p,\bm{\alpha}}(x) = \frac{1}{\sqrt{2\pi\beta}} e^{ i (\frac{2\pi n}{\beta}\tau + m\theta + pz)}J_{m}(k r )\tau_{\bm{\alpha}}\,,
   % \end{align}
    %\begin{align}
%\lambda_{n,m,k,p,\bm{\alpha}}=\qty(\frac{2\pi {n} + \bm{\phi}\!\cdot\!{\bm{\alpha}}}{\beta} +  i \omega{m} )^2 + {k}^2 + {p}^2\,\\
    %\qty(n,m\in\mathbb{Z}\,,\ k\in\mathbb{R}^+\,,\ p\in\mathbb{R}\,,\bm{\alpha}\in\Phi)\,.
    %\end{align}
%\end{subequations}
%where $\Phi$ denotes the union of the $\mathfrak{su}(N)$ root system and the zero roots.\par

The contribution from the gauge field requires the vector Laplacian, $-D^2_{\mathrm{v}}$.  For the explicit form of $-D^2_{\mathrm{v}}$, see Ref.~\cite{Chen:2022smf}.
%\begin{equation}
%    (- D^2_{\mathrm{v}})_\mu^{\ \nu} = 
%    \begin{pmatrix}
%    - {D^2_{\mathrm{s}}} + r^{-2} & 2r^{-3}\partial_\theta & 0 & 0 \\
%    -2r^{-1}\partial_\theta & -  r  {D^2_{\mathrm{s}}}r^{-1} + r^{-2} & 0 & 0 \\
%    0 & 0 & - {D^2_{\mathrm{s}}} & 0 \\
%    -2 r^{-1} \iomega \partial_\theta & 2 r^{-1} \iomega\partial_ r  & 0 & - {D^2_{\mathrm{s}}}
%    \end{pmatrix}\, . \label{Dvector}
%\end{equation}
%\begin{align}
    %\Xi_{n,m,k,p,\bm{\alpha}}^{(1)}(x) &= \Psi_{n,m,k,p,\bm{\alpha}}(x)
    %\begin{pmatrix}
    %1 \\ 0 \\ 0 \\ 0
    %\end{pmatrix}\,,\\
    %\Xi_{n,m,k,p,\bm{\alpha}}^{(2)}(x) &= \Psi_{n,m,k,p,\bm{\alpha}}(x)
    %\begin{pmatrix}
    %0 \\ 0 \\ 0 \\ 1
    %\end{pmatrix}\,.
%\end{align}
%The two transverse eigenmodes have nontrivial tensorial structure:
%\begin{align}
    %\Xi_{n,m,k,p,\bm{\alpha}}^{(\pm)}(x) &= \frac{1}{2\sqrt{\pi\beta}} e^{ i (\frac{2\pi n}{\beta}\tau + m\theta + pz)}\\
    %& \ \ \times J_{m\pm1}(k r)\tau_{\bm{\alpha}}
    %\begin{pmatrix}
    %- i \omega r  \\  r  \\ \pm i  \\ 0
    %\end{pmatrix}\,.
%\end{align}
The eigenvalue spectrum is the same as the scalar one.  Each eigenmode has four polarization degrees of freedom, and two out of four are canceled by the ghost contribution.  After some calculations, we arrive at the Polyakov loop potential resulting from the two physical (transverse) modes as
\begin{align}
    V_g(\bphi;\tiomega) = \frac{T}{4\pi^2} \sum_{\balpha\in\Phi}
    \sum_{l\in\mathbb{Z}}
    & \int_0^\infty \!\! \kp \rmd{\kp}
    \int_{-\infty}^{+\infty} \!\!\! \rmd{k_z} \notag\\
    & \times \Bigl[ J^2_{l-1}(\kp r) + J^2_{l+1}(\kp r) \Bigr] \;
    \mathrm{Re} \ln\qty(1- \rme^{  \rmi \bm{\phi}\cdot\bm{\alpha} - \rmi\tiomega l - \beta|\bm{k}|})\,.
\end{align}
For notational brevity, we introduced a dimensionless imaginary angular velocity; $\tiomega=\iomega/T$.  By expanding the logarithm, we can perform the momentum integration to simplify the above form into
\begin{align}
    V_g(\bphi;\tiomega) = -\frac{2T^4}{\pi^2}
    \sum_{\balpha\in\Phi} \sum_{n=1}^\infty
    \frac{\cos(n\bphi\cdot\balpha)\cos(n\tiomega)}{\Bigl\{n^2 + 2\tilde{r}^2\bigl[1-\cos(n\tiomega)\bigr]\Bigr\}^2} \,.
    \label{eq:poten_g}
\end{align}
The potential is dependent on the dimensionless radial distance; $\tilde{r}=r T$.  The potential is minimized at the optimal value of $\bphi$, and the Polyakov loop expectation value, $L(\bphi)$, is evaluated accordingly.  The denominator has singularities at $\tilde{r}\neq 0$ even for $\bphi=0$ (i.e., free theory), which is consistent with Ref.~\cite{Chernodub:2022qlz}.

Specifically, for the $SU(3)$ Yang-Mills theory, the background gauge field is $A_{\text{B}4} = (\phi_1 T^3 + \phi_2 T^8)/g\beta$, where $T^3$ and $T^8$ constitute the Cartan subalgebra of $\mathfrak{su}(3)$.  The traced Polyakov loop in the fundamental representation of $SU(3)$ is
\begin{equation}
    |L| = \frac{1}{3} \biggl|\tr \exp\biggl(\rmi g\int_0^\beta A_{\text{B}4}\, d\tau\biggr)\biggr|
    = \frac{1}{3}\sqrt{4\cos^2\Bigl(\frac{\phi_1}{2}\Bigr) + 4\cos\Bigl(\frac{\phi_1}{2}\Bigr)\cos\Bigl(\frac{\sqrt{3}\phi_2}{2}\Bigr) + 1}\, .
\end{equation}
In our previous work~\cite{Chen:2022smf}, we showed that confinement, $|L|=0$, is realized for $\tiomega \ge \pi/2$ at $\tilde{r}=0$.

%--- figure ---%
\begin{figure}
    \centering
    \includegraphics[width=0.485\columnwidth]{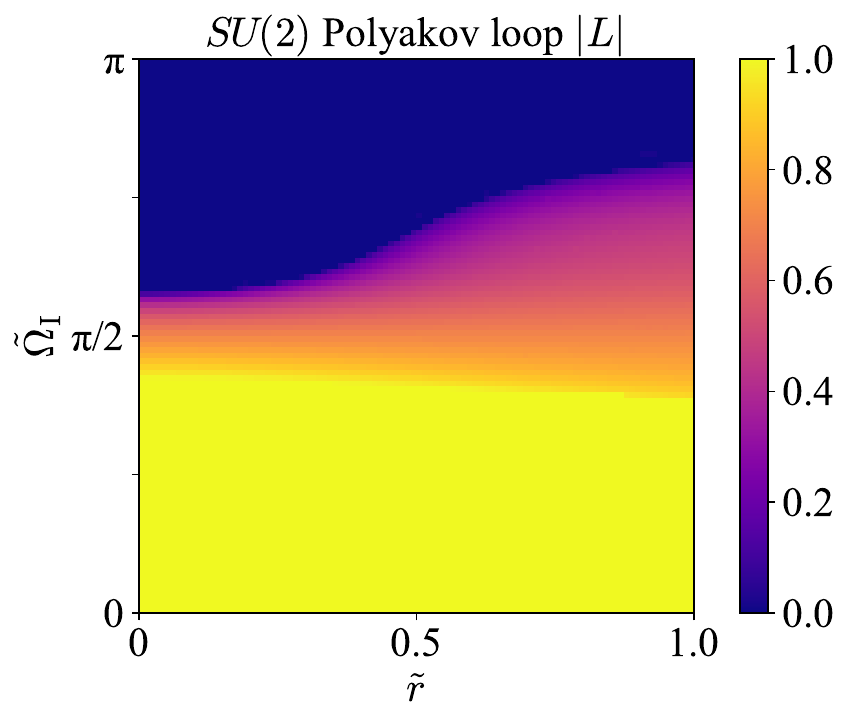}
    \hspace{0.2em}
    \includegraphics[width=0.485\columnwidth]{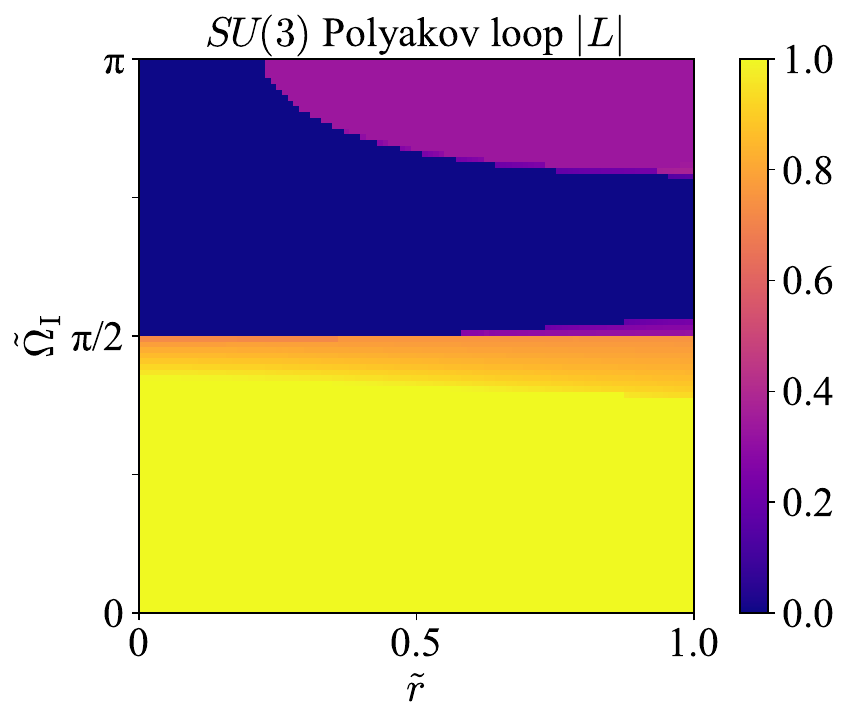}
    \caption{(Left) Polyakov loop $|L|$ for $SU(2)$ as a function of dimensionless radial distance $\tilde{r}$ and dimensionless imaginary angular velocity $\tiomega$. (Right) The same plot of $|L|$ for $SU(3)$.  The confined phase with $|L|=0$ is bounded by the first-order phase transitions.}
    \label{fig:inhomoPLP}
\end{figure}
%--- figure ---%

It is a straightforward exercise to find the global minima of the potential~\eqref{eq:poten_g} for $\tilde{r}\neq 0$.  Figure~\ref{fig:inhomoPLP} shows the results from such extensive analyses of Eq.~\eqref{eq:poten_g} for $SU(2)$ (left) and $SU(3)$ (right).  It is notable that both cases generally develop $r$-dependent spatial structures.  For the $SU(2)$ case as shown in the left of Fig.~\ref{fig:inhomoPLP}, the Polyakov loop changes to zero, indicating confinement, with second-order phase transition as $\tiomega$ grows up.  In the previous paper~\cite{Chen:2022smf}, we focused on two edges of $\tilde{r}=0$ and $\tiomega=\pi$ only.  Our present results imply that, for $\tiomega\simeq 3\pi/4$ for example, there should appear a spatial interface separating the confined phase for $\tilde{r}\lesssim 0.5$ and the deconfined phase for $\tilde{r}\gtrsim 0.5$.  We can confirm a qualitatively similar trend for the $SU(3)$ case, but the detailed shape looks different as shown in the right of Fig.~\ref{fig:inhomoPLP}.  In this case of $SU(3)$, the homogeneously confined phase is realized around $\tiomega \lesssim 3\pi/4$, and the interface between confinement and deconfinement emerges only when $\tiomega$ becomes larger.

We should emphasize that this $\tilde{r}$ dependence of inhomogeneity is qualitatively consistent with the lattice-QCD results~\cite{Braguta:2023iyx}.  As closely discussed in our previous work~\cite{Chen:2022smf}, our $\tiomega$ dependence of the Polyakov loop is opposite to the lattice-QCD case; $|L|$ goes smaller with larger \textit{imaginary} rotation in our perturbative calculation, while $|L|$ goes smaller with larger \textit{real} rotation in the lattice-QCD simulation.  However, in the both cases of ours and the lattice-QCD calculations, $|L|$ becomes larger (smaller) with larger $\tilde{r}$ for a given imaginary (real) angular velocity.  This makes a sharp contrast to preceding works~\cite{Chen:2015hfc,Braga:2022yfe,Yadav:2022qcl,Chernodub:2020qah,Chernodub:2022veq} and might be a key observation to resolve the controversy of rotating QCD matter.

%--- figure ---%
\begin{figure}
    \centering
    \includegraphics[width=0.43\columnwidth]{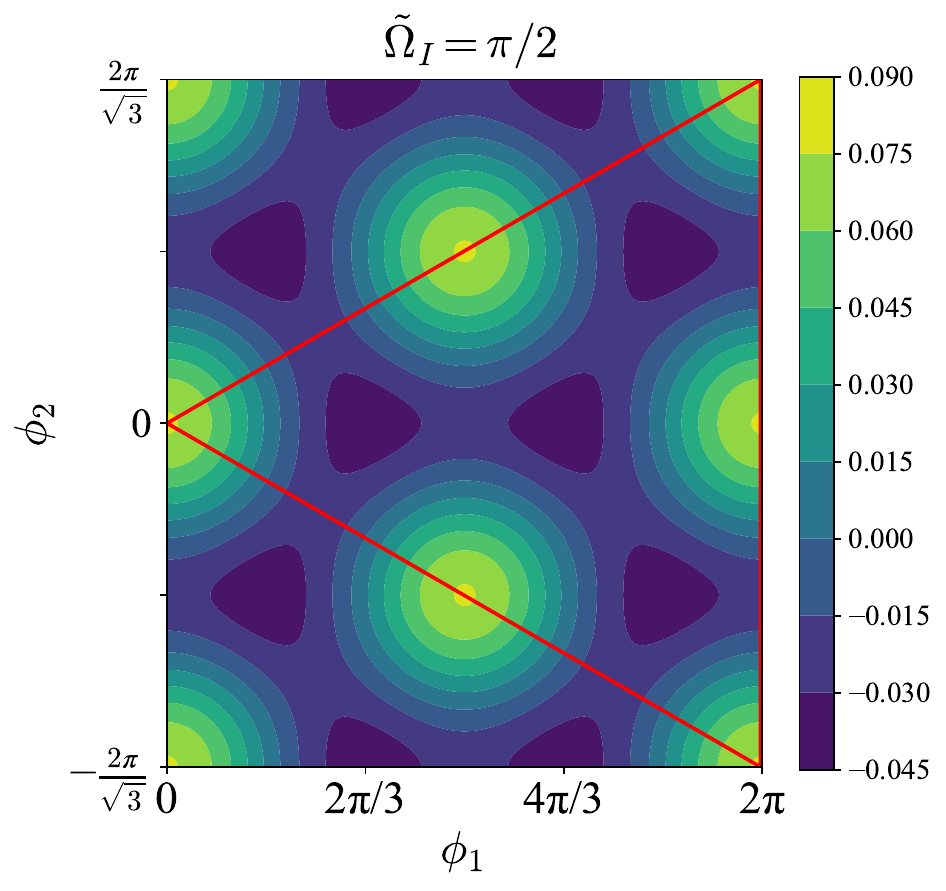}
    \hspace{1em}
    \includegraphics[width=0.43\columnwidth]{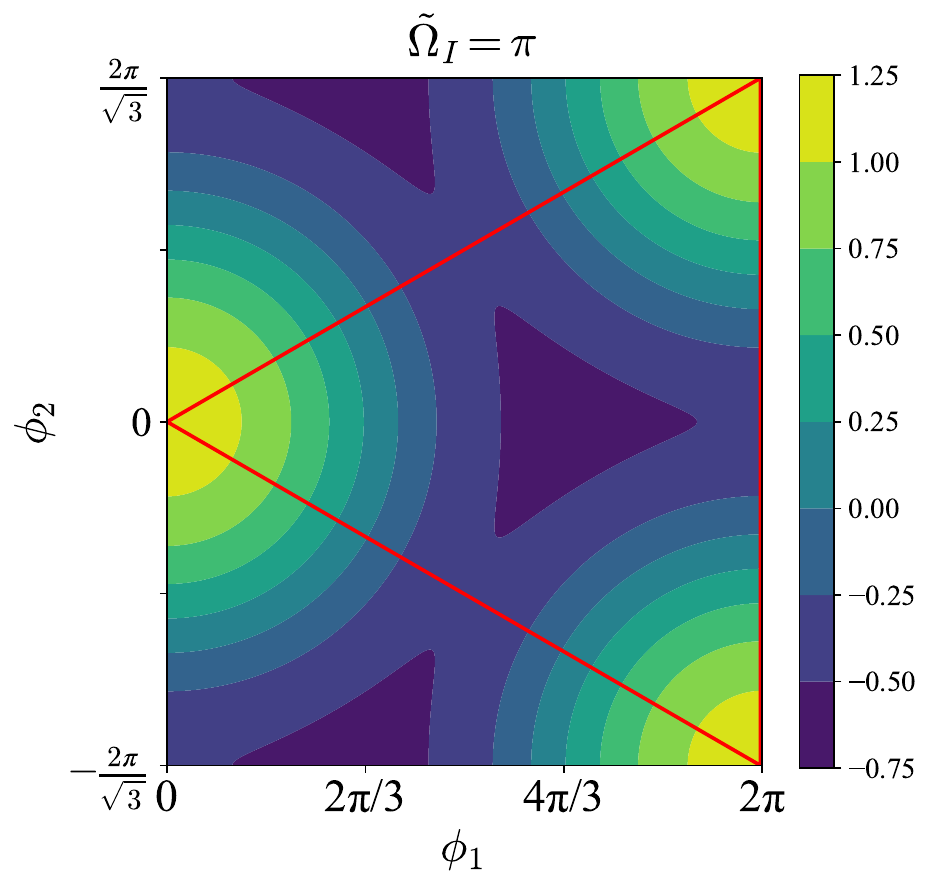}
    \caption{$SU(3)$ effective potential as a function of $\phi_1$ and $\phi_2$ for $\tiomega=\pi/2$ (left) and $\tiomega=\pi$ (right) with $\tilde{r}=0$ fixed.  The dark (light) colored region has smaller (larger) potential values.  The triangular domain indicated by the red line is sufficient for the minimum search.}
    \label{fig:emergentZ2_wis}
\end{figure}
%--- figure ---%

Now, let us turn our attention to the Polyakov loop potential and the symmetry properties.  Since we sum up all the roots in $\Phi$, different $(\phi_1, \phi_2)$ pairs can have the same potential and the Polyakov loop value according to Weyl symmetry of the $SU(3)$ root lattice.  We can see the characteristic patterns of the potential minima in Fig.~\ref{fig:emergentZ2_wis}.  The repetition of the minima reflects Weyl symmetry.  The red triangle region with three edges, $(\phi_1, \phi_2)=(0,0)$, $(2\pi, 2\pi/\sqrt{3})$, $(2\pi, -2\pi/\sqrt{3})$, is the fundamental domain and we can identify the state of matter from the minimum inside this triangular domain.  The triangle has $S_6$ geometric symmetry, including three-fold rotational symmetry, which manifests center symmetry, and two-fold reflective symmetry, which manifests charge conjugation. The potential shape visualized by the pattern in Fig.~\ref{fig:emergentZ2_wis} does not change by $2\pi/3$ rotation but the phase of the Polyakov loop, $L$, does.  The center of the triangle at $(4\pi/3, 0)$ corresponds to $L=0$, that is, a center symmetric vacuum.  Although the potential minima may break center symmetry, the charge conjugation symmetry is never broken. 

We point out a nontrivial observation in Fig.~\ref{fig:emergentZ2_wis};  an emergent symmetry is realized at $\tiomega = \pi/2$.  We observe a reflective mirror on the line of $\phi_1=\pi$.  
% We note that charge conjugation, e.g., $\phi_2\leftrightarrow -\phi_2$, always exists, but the emergent symmetry $\phi_1-\pi\leftrightarrow -(\phi_1-\pi)$ is exceptionally present only for $\tiomega = \pi/2$. 
This emergent symmetry comes from the vanishing of odd-$n$ terms in the one-loop potential~\eqref{eq:poten_g} at $\tiomega = \pi/2$ and exists for not only $\tilde{r}=0$ but any radius.  It could be either a one-loop artifact or a genuine symmetry.  In the latter case, it has to be a non-invertible symmetry like that in the 2D critical Ising model since it exchanges the unbroken and broken vacua.

%--- figure ---%
\begin{figure}
    \centering
    \includegraphics[width=0.43\columnwidth]{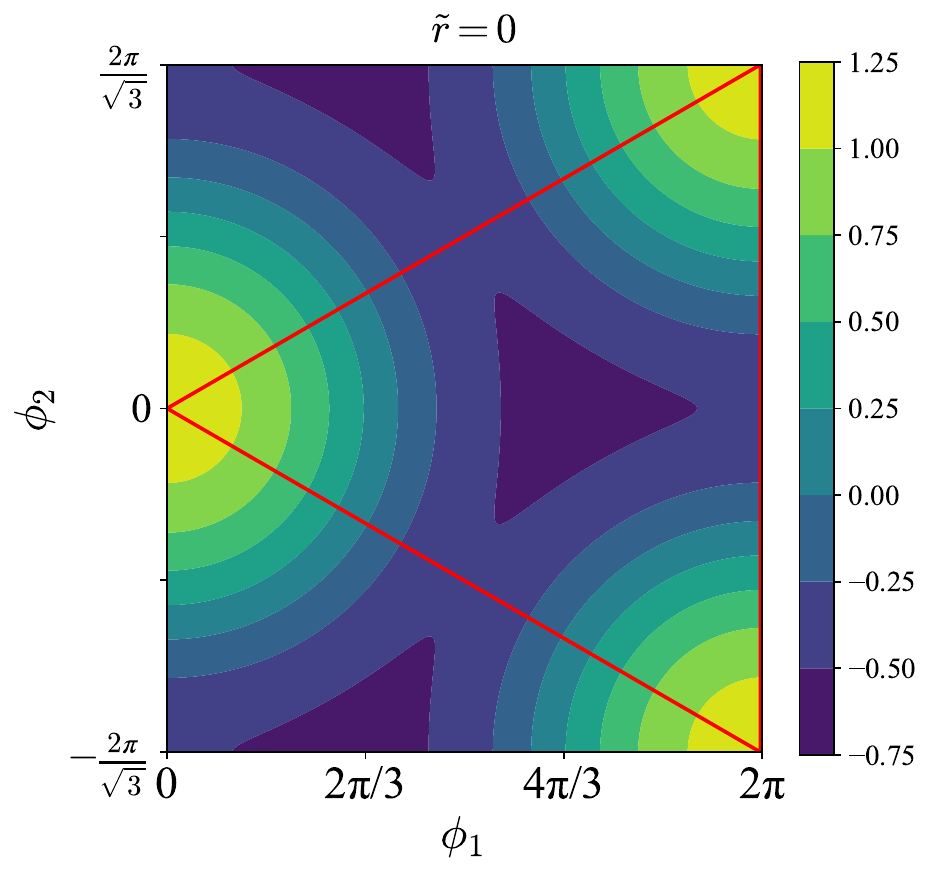}
    \hspace{1em}
    \includegraphics[width=0.43\columnwidth]{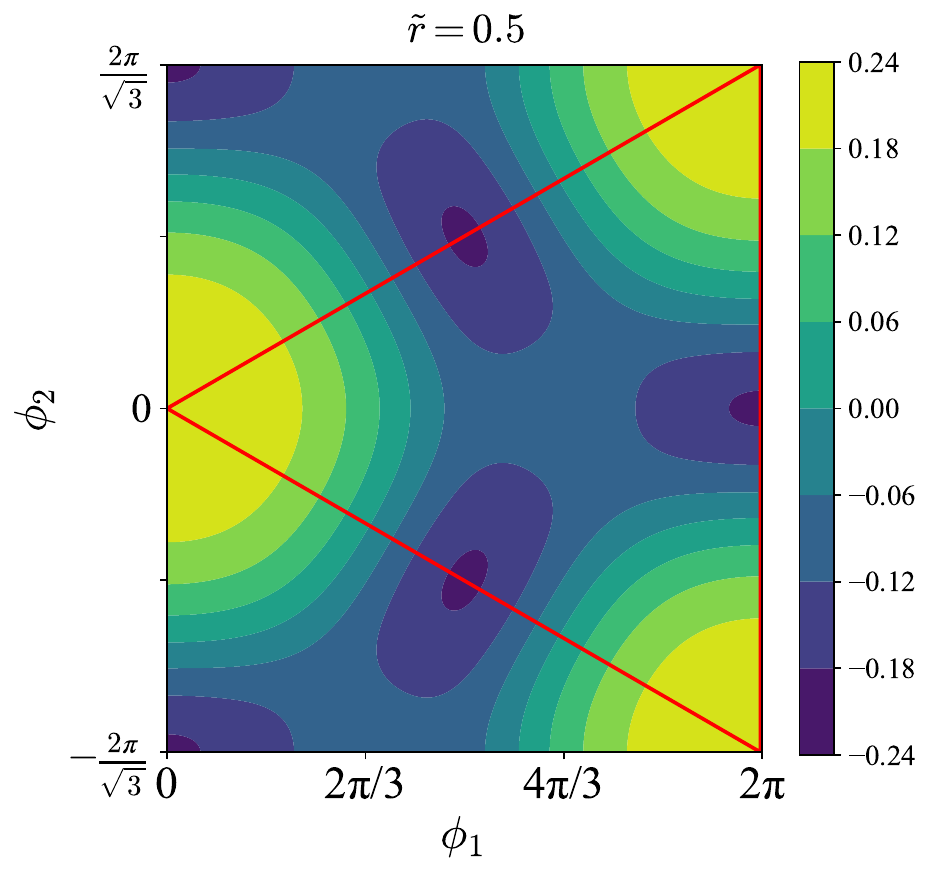}
    \caption{$SU(3)$ effective potential for $\tilde{r}=0$ (left) and $\tilde{r}=0.5$ (right) with $\tiomega=\pi$ fixed.}
    \label{fig:emergentZ2_rs}
\end{figure}
%--- figure ---%

In this work, we also quantify the spatial inhomogeneity.  In Fig.~\ref{fig:emergentZ2_rs}, we plot the Polyakov loop potential for $\tilde{r}=0$ (confined phase) and $\tilde{r}=0.5$ (deconfined phase) at $\tiomega=\pi$ in the $SU(3)$ case.  It is intriguing that the potential minima are located in a different way from the ordinary perturbative vacuum where three vertices of the triangle minimize the potential.  In this sense, this deconfined phase discovered in the upper right region in the right panel of Fig.~\ref{fig:inhomoPLP} may have exotic properties different from the ordinary one.

%% ------------- Tolman--Ehrenfest effect ------------- %%
%Let us discuss about this discrepancy.
%The matter is the definition of the temperature.
%Previous studies considered the Tolman--Ehrenfest (TE) temperature~\cite{Tolman:1930zza,Tolman:1930ona}.
%This is the local temperature, which is defined in a curved space-time by the equation: $T_{\mathrm{TE}}(r) \sqrt{g_{44}} = \mathrm{const.}$.
%In the imaginary rotating case, $T_{\mathrm{TE}}(r)$ is obtained by $ T_{\mathrm{TE}}(r) = T_{\mathrm{TE}}(0)/\sqrt{1+r^2\iomega^2}$ so that this local temperature decreases by $r$.
%The idea of the previous studies is then even at high $T_{\mathrm{TE}}(0)$, $T_{\mathrm{TE}}(r)$ can be lower than the critical temperature.
%While the center is deconfined, the outside can be confined.
%This appears to be correct, but we should note that 
%% the TE temperature is only an apparent temperature.
%the $r$ dependence can also appear to the critical temperature $T_c$.
%So it should not only TE effect that determine the confinement-deconfinement phase transition.
%This point was missed by previous studies.\par

%% ------------- Tolman--Ehrenfest effect ------------- %%

We mention the difference from the previous arguments~\cite{Chernodub:2020qah} based on the Tolman-Ehrenfest (TE) effect.
In our calculation, we treat $T$ as a Lagrange multiplier to conserve the total energy.  By construction, $T$ is a global variable without $r$ dependence.  Therefore, in our study, we do not need to introduce the apparent temperature as a result of the TE effect.  Nevertheless, the calculations naturally lead to $r$ dependent structures.
%and equals $T_{\mathrm{TE}}(0)$.
%Nevertheless, our result showed the $r$ dependence on the critical temperature.
%This then must be the pure effect of rotation on the confinement-deconfinement phase transition.
%Note that, since our study compared the potential value against $\phi$ at the each $r$, the definition of the temperature do not change the result.
%In addition, since the dimension-less potential $V_g/T^4$ has no $T$ dependence any more, our result can be used for any high temperature: the center of rapidly imaginary rotating systems is always confined even at many times the critical temperature of static matter.
%This agrees with the latest lattice calculations~\cite{Braguta:2023iyx}, which showed that the center of the system do not become deconfined.
%This means the critical temperature can be infinite for such systems.

%%%%%%%%%%
\section{Chiral symmetry breaking in the perturbatively confined phase}

We can repeat similar calculations including dynamical quark contributions that break center symmetry explicitly.  We can also address a relation between confinement and chiral symmetry breaking from two (approximate) order parameters, namely, the Polyakov loop and the dynamical quark mass, $m$, which is rooted in the chiral condensate.  In the perturbative treatment, the pressure (the free energy) is maximized (minimized) for $m=0$, and the dynamical mass generation is energetically disfavored.  It is quite interesting what would happen in the perturbatively confined phase with imaginary rotation.

The theoretical treatments of fermions in the rotating frame are found in Refs.~\cite{Chen:2015hfc,Ebihara:2016fwa,Jiang:2016wvv}.  We should be careful about the fact that the Polyakov loop coupling with quarks is given by the fundamental representation.  
%Let us review the calculation of the fermions' effective potential.
We can calculate the fermionic partition function by imposing an aperiodic thermal boundary condition or equivalently considering the ordinary anti-periodic boundary condition in the rotating frame.
In this paper, for fermions, we choose the latter.
It should be noted that the fermion interactions appear from gauge fluctuations which are beyond the one-loop perturbative order.  In this way, we can locate the onset of instability toward chiral symmetry breaking, but finding the physical value of $m\neq 0$ needs non-perturbative interactions that we do not include in the present study.
The fermionic partition function is the determinant of the Dirac operator $\gamma^\mu G_{{\rm B}\,\mu} + m$ in the rotating frame, i.e.,
\begin{equation}
    \mathcal{Z}_{{\rm f} T, \omega} = \Det (\gamma^\mu G_{{\rm B}\,\mu} + m) \,.
\end{equation}
Here, $G_{{\rm B}\,\mu} = D_{\mu} - \Gamma_\mu$ is the covariant derivative including the $A_{{\rm B}4}$ background field with $\Gamma_\mu = -\frac{i}{4}\sigma^{ij} \, \omega_{\mu ij}$, where $\sigma^{ij} = \frac{i}{2} [\hamma^i, \hamma^j]$ and $\omega_{\mu ij} = g_{\rho \sigma} \, e_i^{\ \rho} \, \qty(\partial_\mu e^{\ \sigma}_j + \Gamma^\sigma_{\mu\nu}\, e^{\ \nu}_j)$.
We denote the gamma matrices of the flat space-time by $\hamma^i$ and $\gamma^\mu = e_i^{\ \mu} \hamma^i$.
After all, we arrive at the expression for the Polyakov loop potential per one fermion (particle or anti-particle but without flavor degrees of freedom) as
\begin{align}
    V_f(\bphi;\tiomega) 
    &= -\frac{T}{4\pi^2} \sum_{\bm{\mu}\in\Phi_f} \sum_{l\in\mathbb{Z}} \int_0^\infty \kp\, \rmd\kp \int_{-\infty}^\infty \rmd k_z \notag\\
    &\qquad\times \Bigl[ J_{l}^2(\kp r) + J_{l+1}^2(\kp r) \Bigr] \Re \ln \Bigl( 1 + \rme^{\rmi\bphi\cdot\bm{\mu} - \rmi\tiomega(l+1/2) - \beta\sqrt{\bm{k}^2 + m^2}} \Bigr) \,,
    \label{eq:perturbative_rotating_potential}
\end{align}
where $\Phi_f$ denotes the set of the fundamental weights of $\mathfrak{su}(\Nc)$.
The limit of $\bphi = 0$ renders the above expression to the free quark and anti-quark gas energy with imaginary rotation derived from statistical mechanics.

First, we shall consider the interplay of $|L|$ and $m$ at $\tilde{r} = 0$ by setting $\Nc = \Nf = 2$.  Then, the total effective potential is given by $V_g + 2\Nf V_f$ (where $2$ comes from the sum over particle and anti-particle).  In our numerical calculation, we consider $\phi \in [0, 2\pi]$ and $\tilde{m}=m/T \in [0, 10]$.  We note that the former region for $\phi$ is sufficient thanks to $SU(2)$ Weyl symmetry.  Because we do not model quark interactions to keep our analysis as model independent as possible, $m$ blows up in the phase where chiral symmetry is fully broken.  We thus set a numerical cutoff on $\tilde{m}$ in the practical calculation.

%--- figure ---%
\begin{figure}
    \centering
    \includegraphics[width=0.43\columnwidth]{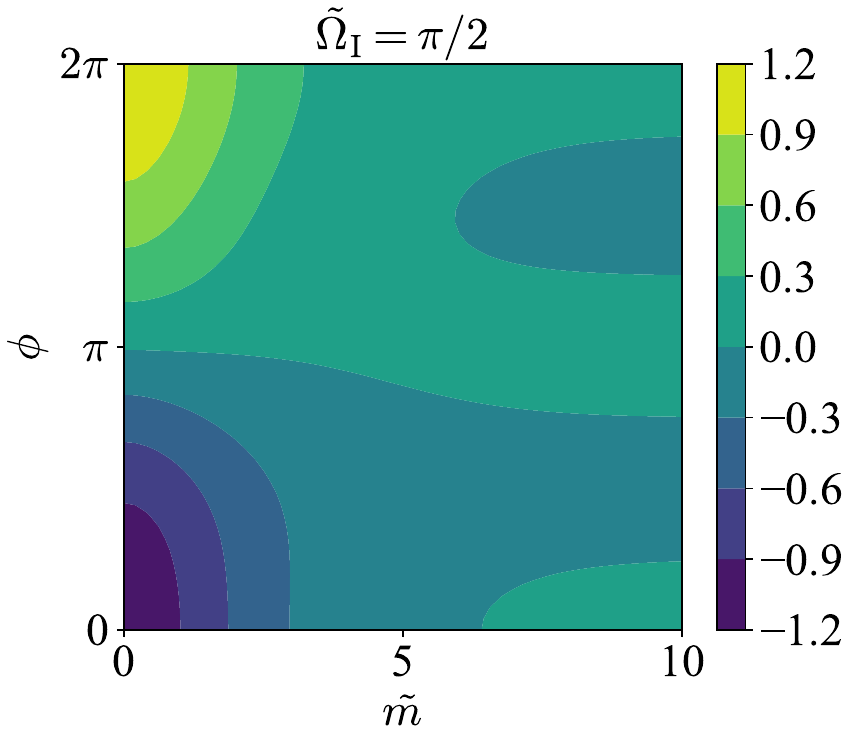}
    \hspace{1em}
    \includegraphics[width=0.43\columnwidth]{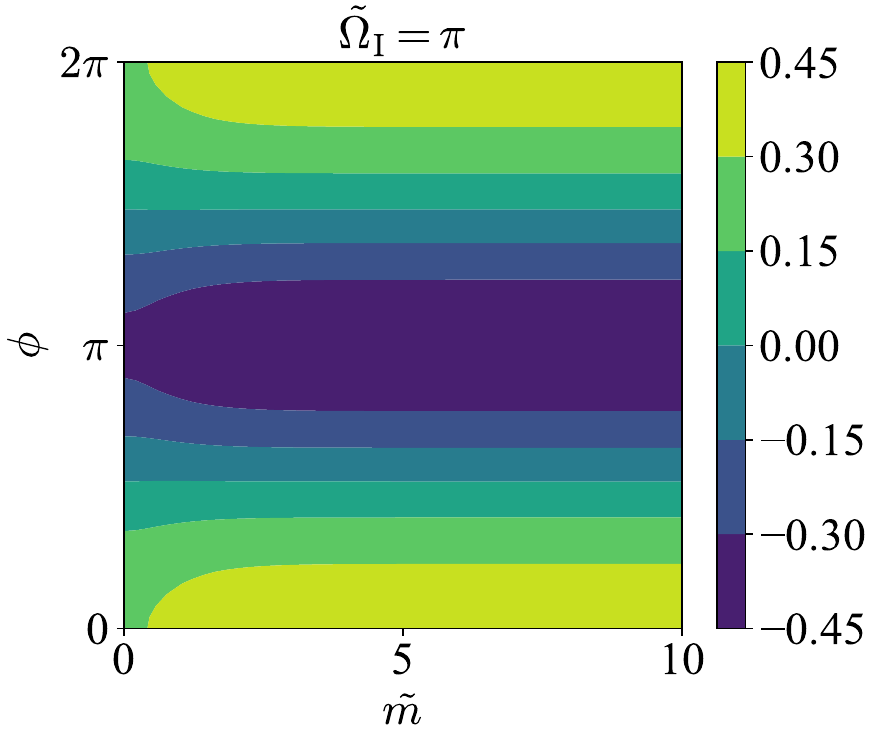}
    \caption{Potential contour as functions of $\phi$ (Polyakov loop) and $\tilde{m}$ (dimensionless dynamical mass) for $\Nc = \Nf = 2$, $\tiomega=\pi/2$ (left) and $\tiomega=\pi$ (right). 
 The darker (lighter) color indicates the smaller (larger) value and the potential minimum has the darkest color.}
    \label{fig:SU2phimplot}
\end{figure}
%--- figure ---%

Figure~\ref{fig:SU2phimplot} shows the potential contour as functions of two order parameters in the $SU(2)$ case.  The minimum of the potential is given by the darkest spot in the figures.  In the left of Fig.~\ref{fig:SU2phimplot} for $\tiomega=\pi/2$, the minimum is still located at $\tilde{m}=0$ and $\phi=0$ (i.e., $L=1$).  This is different from the pure gluonic case in the left of Fig.~\ref{fig:inhomoPLP};  fermions generally favor deconfinement.  Interestingly in the right of Fig.~\ref{fig:SU2phimplot} for $\tiomega=\pi$, however, the minimum moves to $\phi\simeq\pi$ and $\tilde{m}$ blows up to the cutoff value, which signifies confinement and spontaneous chiral symmetry breaking.

%From the figure, $m$ should be larger than 2 GeV at $\iomega/T = \pi$.
%However, we only need a finite $m$ to see the chiral symmetry breaking so this is sufficient.
%Nevertheless the obtained $\phi$ there cannot be the same as the true $\phi$ that is obtained by true $m$, the figure shows that favored $\phi$ should be $\pi$ with any true $m$ would be obtained.\par

%--- figure ---%
\begin{figure}
    \centering
    \includegraphics[width=0.45\columnwidth]{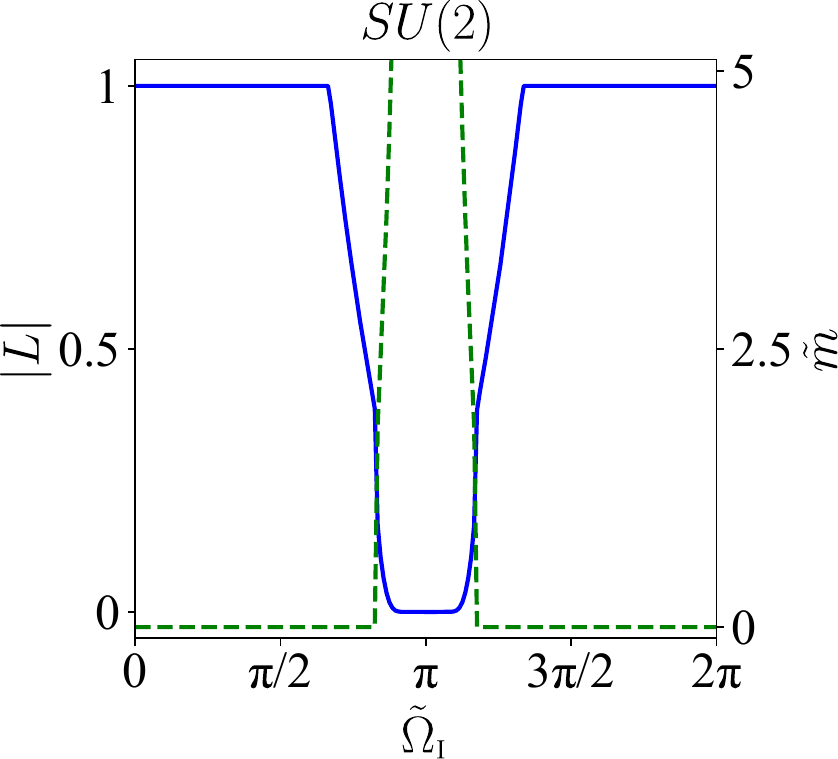}
    \hspace{1em}
    \includegraphics[width=0.45\columnwidth]{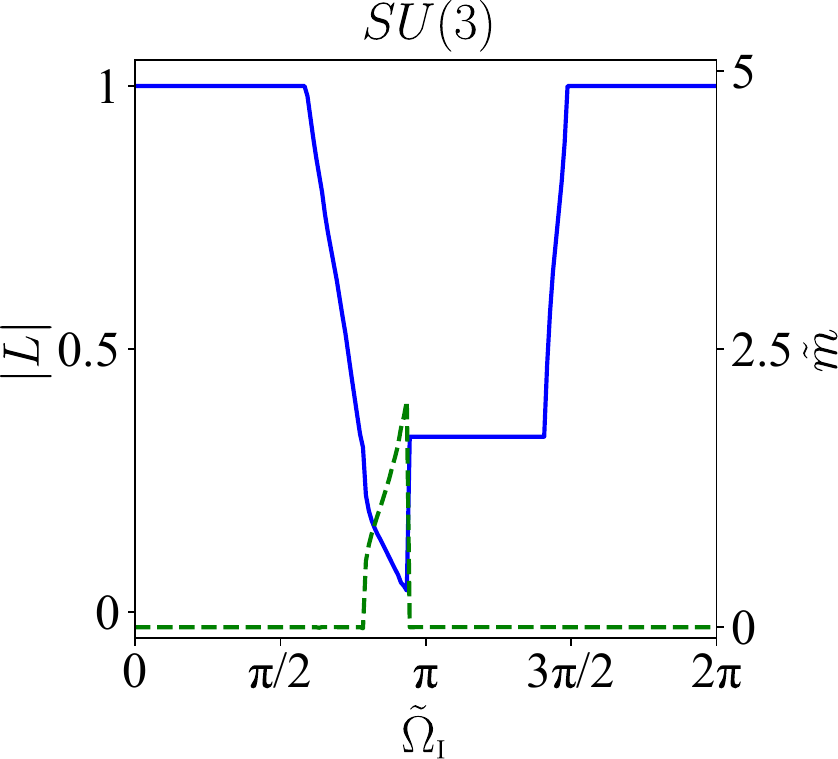}
    \caption{Order parameters, $|L|$ and $\tilde{m}$, as functions of $\tiomega$ for $\Nf=2$, $\Nc=2$ (left) and $\Nc=3$ (right). The blue solid line represents the Polyakov loop $|L|$, and the green dashed line represents the dynamical mass $\tilde{m}$. The horizontal axis is discretized by $200$ points with equal spacing to capture spiky structures.}
    \label{fig:SU2PM}
\end{figure}
%--- figure ---%

We make the plots for the behavior of the Polyakov loop $|L|$ and the dynamical mass $\tilde{m}$ in Fig.~\ref{fig:SU2PM}.  The blue solid line represents $|L|$, while the green dashed line represents $\tilde{m}$.  With increasing $\tiomega$, the Polyakov loop goes smaller and the dynamical mass grows up.  Then, a confined and chiral symmetry broken state is favored near $\tiomega\simeq \pi$.  For the $SU(2)$ case as shown in the left of Fig.~\ref{fig:SU2PM}, the behavior is $2\pi$ periodic because the $SU(2)$ gauge group covers the fermionic parity $(-)^F$.
% because the potential satisfies $V_f(\phi, \tiomega + 2\pi) = V_f(\phi + 2\pi, \tiomega)$. 
Also we see that the behavior is symmetric around $\tiomega=\pi$.  For the $SU(3)$ case in the right of Fig.~\ref{fig:SU2PM}, in contrast, the order parameters depend on $\tiomega$ in a complicated way and the confined phase seems to be less favored.
Also the periodicity becomes $4\pi$ because the $SU(3)$ gauge group does not contain $(-)^F$.
%Note that fermionic matter is also included here but the gluonic matter is dominant around $\iomega/T = \pi$ and so we have exactly zero Polyakov loop.\par

%--- figure ---%
\begin{figure}
    \centering
    \includegraphics[width=0.485\columnwidth]{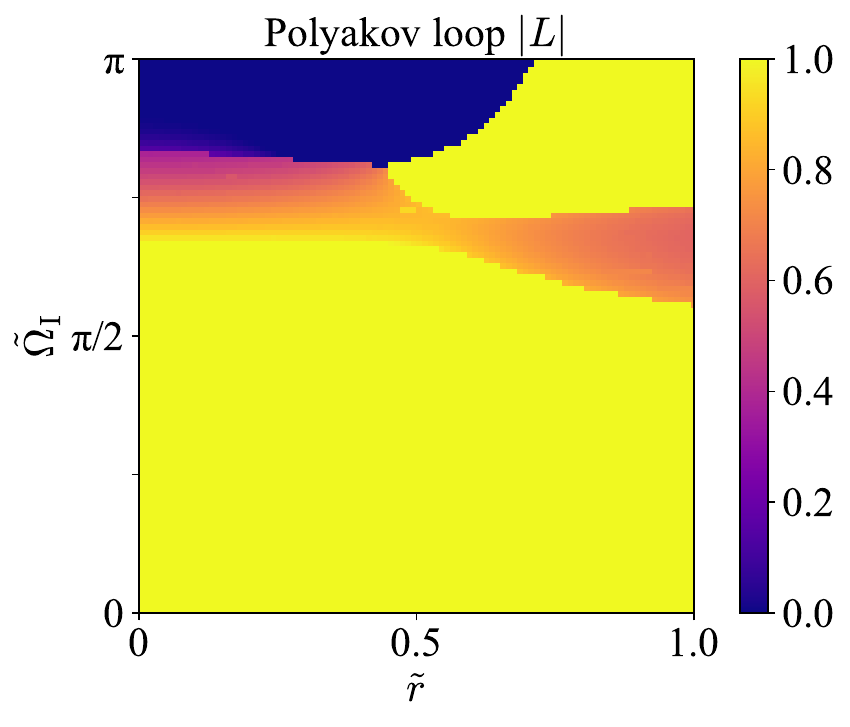}
    \hspace{0.2em}
    \includegraphics[width=0.485\columnwidth]{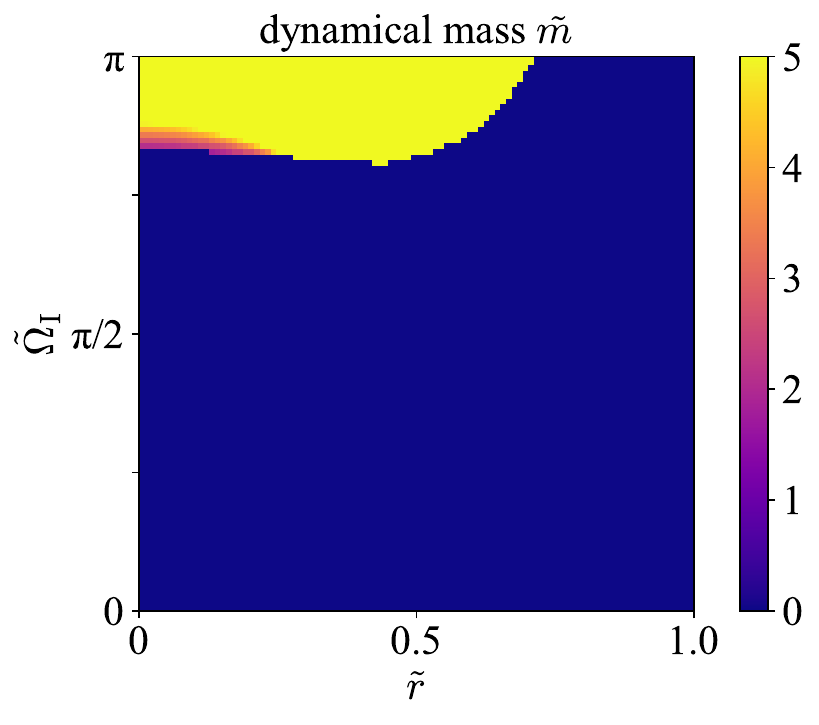}
    \caption{Order parameters in the $SU(2)$ case; the Polyakov loop (left) and the dynamical mass (right) as functions of dimensionless radial distance $\tilde{r}$ and dimensionless imaginary angular velocity $\tiomega$. To reduce the numerical cost, the mesh resolution is chosen to be $100\times 100$.}
    \label{fig:su2_full}
\end{figure}
%--- figure ---%

In the same way as the pure gluonic theory, we have quantified the inhomogeneous structures in Fig.~\ref{fig:su2_full}.  The left and the right panels show the Polyakov loop and the dynamical mass, respectively, as functions of $\tilde{r}$ and $\tiomega$.  We have performed this calculation for the $SU(2)$ case only.  As compared to the left of Fig.~\ref{fig:inhomoPLP} in the pure gluonic case, the perturbatively confined region is shrunk to the upper left corner.  Interestingly, in view of the right of Fig.~\ref{fig:su2_full}, confinement and chiral symmetry breaking are locked together and the phase transitions occur simultaneously even at large imaginary rotation.
%In addition, there appear some substructures, but the numerical convergence seems to be insufficient.  We have tried, but the potential shape is quite complicated and it is difficult to obtain smooth results in the whole region at $\tilde{r}\gtrsim 0.5$.
It would be an interesting future problem to probe a possibility of fractal structures for $\tiomega/(2\pi)$ given by a rational number, which is not visible in the present analysis.
%This complication might be a precursory of possible fractals~\cite{Ambrus:2023bid}.  More investigations about these substructures as well as the full $SU(3)$ calculations are left as future problems.

%--- figure ---%
%\begin{figure}[tb]
    %\centering
    %\includegraphics[width = \linewidth]{fig_SU2Mpot}
    %\caption{Caption}
    %\label{fig:SU2Mpot}
%\end{figure}
%--- figure ---%

%Finally let us add explanation about the constituent mass.
%To determine the constituent mass, we should note that the vacuum part of the free energy $V_0$ also has the $m$ dependence.
%We should consider both (vacuum and thermodynamic) part of the free energy.
%Because chiral symmetry is broken at vacuum, $V_0(m)$ has a minimum at large $m$ as the dotted line in FIG.~\ref{fig:SU2Mpot}.
%The thermodynamic part shifts this minimum as represented by dash-dotted line.
%If the system is not rotating, total free energy is minimized by zero dynamical mass and so the chiral symmetry is restored.
%As rotating velocity increase, the potential is flipped and finite mass minimize the potential.
%Therefore the chiral symmetry is broken in rapidly rotating systems.
%This behavior coincides with well-known behavior of low temperature systems.
%The system is confined and chiral symmetry of the system is broken.
%As we found in our letter~\cite{Chen:2022smf}, rotating system has the similar character to the hadronic phase.\par

%%%%%%%%%%
\section{Conclusions}

We considered the inhomogeneous structures and the chiral symmetry breaking in the perturbatively confined phase which is realized by rotation with imaginary angular velocity, $\tiomega$.  In both color $SU(2)$ and $SU(3)$ cases, the perturbatively confined phase is induced at the rotation center, $\tilde{r}=0$, for $\tiomega$ above a certain threshold.  We found that the perturbative confinement may break down in the $\tilde{r}\neq0$ region away from the rotation center.  This indicates the existence of phase interface between confinement and deconfinement and the critical temperature is also $\tilde{r}$ dependent accordingly.  Our results predict that the critical temperature should decrease with increasing $\tilde{r}$ in the presence of imaginary rotation.  This trend agrees with the latest results from the lattice simulation.
%Since our definition of the temperature do not contain any $r$ dependence, the $r$ dependence in our result should be the pure effect of rotation.
Although the analytical continuation to real rotation has some subtle points, the inhomogeneity we have confirmed from the perturbative effective potential can presumably persist in real rotating systems.

We then added the quark contribution to the Polyakov loop effective potential, which is also a function of the dynamical quark mass, $\tilde{m}$.  Our analysis of the potential minimum search showed that confinement and chiral symmetry breaking are tightly correlated in such unconventional systems with imaginary rotation and even without fermionic interaction.  In the presence of quarks that explicitly break center symmetry, the Polyakov loop is only an approximate order parameter.  Still, in the region where the Polyakov loop is vanishingly small, $\tilde{m}$ departs from zero signifying spontaneous breaking of chiral symmetry.  In our treatment, in fact, $\tilde{m}$ blows up when chiral symmetry is broken, so that large $\tilde{m}$ suppresses center symmetry breaking.  In this way, we established simultaneous realization of confinement and chiral symmetry breaking.  We also found highly complicated substructures in the deconfined regions.  The imaginary-rotating inhomogeneous states with dynamical quarks should deserve further investigations in the future. The full understanding of imaginary-rotating matter should be the indispensable foundation for a more important and challenging problem with real rotation which requires a boundary to satisfy the causality condition. We are making progresses along these lines.

\section*{Acknowledgement}
The authors thank
Maxim~Chernodub and Xu-Guang~Huang
for useful discussions.
This work was supported by Japan Society for the Promotion of Science
(JSPS) KAKENHI Grant Nos.\ 22H01216 (K.F.) and 22H05118 (K.F.) and 21J20877 (S.C.) and JST SPRING, Grant Number JPMJSP2108 (Y.S.).

\bibliography{Z2ANDchiral}

%apsrev4-2.bst 2019-01-14 (MD) hand-edited version of apsrev4-1.bst
%Control: key (0)
%Control: author (72) initials jnrlst
%Control: editor formatted (1) identically to author
%Control: production of article title (-1) disabled
%Control: page (0) single
%Control: year (1) truncated
%Control: production of eprint (0) enabled
\begin{thebibliography}{44}%
\makeatletter
\providecommand \@ifxundefined [1]{%
 \@ifx{#1\undefined}
}%
\providecommand \@ifnum [1]{%
 \ifnum #1\expandafter \@firstoftwo
 \else \expandafter \@secondoftwo
 \fi
}%
\providecommand \@ifx [1]{%
 \ifx #1\expandafter \@firstoftwo
 \else \expandafter \@secondoftwo
 \fi
}%
\providecommand \natexlab [1]{#1}%
\providecommand \enquote  [1]{``#1''}%
\providecommand \bibnamefont  [1]{#1}%
\providecommand \bibfnamefont [1]{#1}%
\providecommand \citenamefont [1]{#1}%
\providecommand \href@noop [0]{\@secondoftwo}%
\providecommand \href [0]{\begingroup \@sanitize@url \@href}%
\providecommand \@href[1]{\@@startlink{#1}\@@href}%
\providecommand \@@href[1]{\endgroup#1\@@endlink}%
\providecommand \@sanitize@url [0]{\catcode `\\12\catcode `\$12\catcode
  `\&12\catcode `\#12\catcode `\^12\catcode `\_12\catcode `\%12\relax}%
\providecommand \@@startlink[1]{}%
\providecommand \@@endlink[0]{}%
\providecommand \url  [0]{\begingroup\@sanitize@url \@url }%
\providecommand \@url [1]{\endgroup\@href {#1}{\urlprefix }}%
\providecommand \urlprefix  [0]{URL }%
\providecommand \Eprint [0]{\href }%
\providecommand \doibase [0]{https://doi.org/}%
\providecommand \selectlanguage [0]{\@gobble}%
\providecommand \bibinfo  [0]{\@secondoftwo}%
\providecommand \bibfield  [0]{\@secondoftwo}%
\providecommand \translation [1]{[#1]}%
\providecommand \BibitemOpen [0]{}%
\providecommand \bibitemStop [0]{}%
\providecommand \bibitemNoStop [0]{.\EOS\space}%
\providecommand \EOS [0]{\spacefactor3000\relax}%
\providecommand \BibitemShut  [1]{\csname bibitem#1\endcsname}%
\let\auto@bib@innerbib\@empty
%</preamble>
\bibitem [{\citenamefont {Svetitsky}(1986)}]{Svetitsky:1985ye}%
  \BibitemOpen
  \bibfield  {author} {\bibinfo {author} {\bibfnamefont {B.}~\bibnamefont
  {Svetitsky}},\ }\href {https://doi.org/10.1016/0370-1573(86)90014-1}
  {\bibfield  {journal} {\bibinfo  {journal} {Phys. Rept.}\ }\textbf {\bibinfo
  {volume} {132}},\ \bibinfo {pages} {1} (\bibinfo {year} {1986})}\BibitemShut
  {NoStop}%
\bibitem [{\citenamefont {Son}\ and\ \citenamefont
  {Stephanov}(2001{\natexlab{a}})}]{Son:2000by}%
  \BibitemOpen
  \bibfield  {author} {\bibinfo {author} {\bibfnamefont {D.~T.}\ \bibnamefont
  {Son}}\ and\ \bibinfo {author} {\bibfnamefont {M.~A.}\ \bibnamefont
  {Stephanov}},\ }\href {https://doi.org/10.1134/1.1378872} {\bibfield
  {journal} {\bibinfo  {journal} {Phys. Atom. Nucl.}\ }\textbf {\bibinfo
  {volume} {64}},\ \bibinfo {pages} {834} (\bibinfo {year}
  {2001}{\natexlab{a}})},\ \Eprint {https://arxiv.org/abs/hep-ph/0011365}
  {arXiv:hep-ph/0011365} \BibitemShut {NoStop}%
\bibitem [{\citenamefont {Son}\ and\ \citenamefont
  {Stephanov}(2001{\natexlab{b}})}]{Son:2000xc}%
  \BibitemOpen
  \bibfield  {author} {\bibinfo {author} {\bibfnamefont {D.~T.}\ \bibnamefont
  {Son}}\ and\ \bibinfo {author} {\bibfnamefont {M.~A.}\ \bibnamefont
  {Stephanov}},\ }\href {https://doi.org/10.1103/PhysRevLett.86.592} {\bibfield
   {journal} {\bibinfo  {journal} {Phys. Rev. Lett.}\ }\textbf {\bibinfo
  {volume} {86}},\ \bibinfo {pages} {592} (\bibinfo {year}
  {2001}{\natexlab{b}})},\ \Eprint {https://arxiv.org/abs/hep-ph/0005225}
  {arXiv:hep-ph/0005225} \BibitemShut {NoStop}%
\bibitem [{\citenamefont {Yamamoto}(2013)}]{Yamamoto:2012bd}%
  \BibitemOpen
  \bibfield  {author} {\bibinfo {author} {\bibfnamefont {A.}~\bibnamefont
  {Yamamoto}},\ }\href {https://doi.org/10.1103/PhysRevLett.110.112001}
  {\bibfield  {journal} {\bibinfo  {journal} {Phys. Rev. Lett.}\ }\textbf
  {\bibinfo {volume} {110}},\ \bibinfo {pages} {112001} (\bibinfo {year}
  {2013})},\ \Eprint {https://arxiv.org/abs/1210.8250} {arXiv:1210.8250
  [hep-lat]} \BibitemShut {NoStop}%
\bibitem [{\citenamefont {Andersen}\ \emph {et~al.}(2016)\citenamefont
  {Andersen}, \citenamefont {Naylor},\ and\ \citenamefont
  {Tranberg}}]{Andersen:2014xxa}%
  \BibitemOpen
  \bibfield  {author} {\bibinfo {author} {\bibfnamefont {J.~O.}\ \bibnamefont
  {Andersen}}, \bibinfo {author} {\bibfnamefont {W.~R.}\ \bibnamefont
  {Naylor}},\ and\ \bibinfo {author} {\bibfnamefont {A.}~\bibnamefont
  {Tranberg}},\ }\href {https://doi.org/10.1103/RevModPhys.88.025001}
  {\bibfield  {journal} {\bibinfo  {journal} {Rev. Mod. Phys.}\ }\textbf
  {\bibinfo {volume} {88}},\ \bibinfo {pages} {025001} (\bibinfo {year}
  {2016})},\ \Eprint {https://arxiv.org/abs/1411.7176} {arXiv:1411.7176
  [hep-ph]} \BibitemShut {NoStop}%
\bibitem [{\citenamefont {Chen}\ \emph {et~al.}(2017)\citenamefont {Chen},
  \citenamefont {Fukushima}, \citenamefont {Huang},\ and\ \citenamefont
  {Mameda}}]{Chen:2017xrj}%
  \BibitemOpen
  \bibfield  {author} {\bibinfo {author} {\bibfnamefont {H.-L.}\ \bibnamefont
  {Chen}}, \bibinfo {author} {\bibfnamefont {K.}~\bibnamefont {Fukushima}},
  \bibinfo {author} {\bibfnamefont {X.-G.}\ \bibnamefont {Huang}},\ and\
  \bibinfo {author} {\bibfnamefont {K.}~\bibnamefont {Mameda}},\ }\href
  {https://doi.org/10.1103/PhysRevD.96.054032} {\bibfield  {journal} {\bibinfo
  {journal} {Phys. Rev. D}\ }\textbf {\bibinfo {volume} {96}},\ \bibinfo
  {pages} {054032} (\bibinfo {year} {2017})},\ \Eprint
  {https://arxiv.org/abs/1707.09130} {arXiv:1707.09130 [hep-ph]} \BibitemShut
  {NoStop}%
\bibitem [{\citenamefont {Jiang}\ and\ \citenamefont
  {Liao}(2016)}]{Jiang:2016wvv}%
  \BibitemOpen
  \bibfield  {author} {\bibinfo {author} {\bibfnamefont {Y.}~\bibnamefont
  {Jiang}}\ and\ \bibinfo {author} {\bibfnamefont {J.}~\bibnamefont {Liao}},\
  }\href {https://doi.org/10.1103/PhysRevLett.117.192302} {\bibfield  {journal}
  {\bibinfo  {journal} {Phys. Rev. Lett.}\ }\textbf {\bibinfo {volume} {117}},\
  \bibinfo {pages} {192302} (\bibinfo {year} {2016})},\ \Eprint
  {https://arxiv.org/abs/1606.03808} {arXiv:1606.03808 [hep-ph]} \BibitemShut
  {NoStop}%
\bibitem [{\citenamefont {Wang}\ \emph {et~al.}(2019)\citenamefont {Wang},
  \citenamefont {Wei}, \citenamefont {Li},\ and\ \citenamefont
  {Huang}}]{Wang:2018sur}%
  \BibitemOpen
  \bibfield  {author} {\bibinfo {author} {\bibfnamefont {X.}~\bibnamefont
  {Wang}}, \bibinfo {author} {\bibfnamefont {M.}~\bibnamefont {Wei}}, \bibinfo
  {author} {\bibfnamefont {Z.}~\bibnamefont {Li}},\ and\ \bibinfo {author}
  {\bibfnamefont {M.}~\bibnamefont {Huang}},\ }\href
  {https://doi.org/10.1103/PhysRevD.99.016018} {\bibfield  {journal} {\bibinfo
  {journal} {Phys. Rev. D}\ }\textbf {\bibinfo {volume} {99}},\ \bibinfo
  {pages} {016018} (\bibinfo {year} {2019})},\ \Eprint
  {https://arxiv.org/abs/1808.01931} {arXiv:1808.01931 [hep-ph]} \BibitemShut
  {NoStop}%
\bibitem [{\citenamefont {Chernodub}\ and\ \citenamefont
  {Gongyo}(2017)}]{Chernodub:2016kxh}%
  \BibitemOpen
  \bibfield  {author} {\bibinfo {author} {\bibfnamefont {M.~N.}\ \bibnamefont
  {Chernodub}}\ and\ \bibinfo {author} {\bibfnamefont {S.}~\bibnamefont
  {Gongyo}},\ }\href {https://doi.org/10.1007/JHEP01(2017)136} {\bibfield
  {journal} {\bibinfo  {journal} {JHEP}\ }\textbf {\bibinfo {volume} {01}},\
  \bibinfo {pages} {136}},\ \Eprint {https://arxiv.org/abs/1611.02598}
  {arXiv:1611.02598 [hep-th]} \BibitemShut {NoStop}%
\bibitem [{\citenamefont {Chen}\ \emph {et~al.}(2023)\citenamefont {Chen},
  \citenamefont {Zhu},\ and\ \citenamefont {Huang}}]{Chen:2023cjt}%
  \BibitemOpen
  \bibfield  {author} {\bibinfo {author} {\bibfnamefont {H.-L.}\ \bibnamefont
  {Chen}}, \bibinfo {author} {\bibfnamefont {Z.-B.}\ \bibnamefont {Zhu}},\ and\
  \bibinfo {author} {\bibfnamefont {X.-G.}\ \bibnamefont {Huang}},\ }\href
  {https://doi.org/10.1103/PhysRevD.108.054006} {\bibfield  {journal} {\bibinfo
   {journal} {Phys. Rev. D}\ }\textbf {\bibinfo {volume} {108}},\ \bibinfo
  {pages} {054006} (\bibinfo {year} {2023})},\ \Eprint
  {https://arxiv.org/abs/2306.08362} {arXiv:2306.08362 [hep-ph]} \BibitemShut
  {NoStop}%
\bibitem [{\citenamefont {Sun}\ \emph {et~al.}(2023)\citenamefont {Sun},
  \citenamefont {Xu},\ and\ \citenamefont {Huang}}]{Sun:2023kuu}%
  \BibitemOpen
  \bibfield  {author} {\bibinfo {author} {\bibfnamefont {F.}~\bibnamefont
  {Sun}}, \bibinfo {author} {\bibfnamefont {K.}~\bibnamefont {Xu}},\ and\
  \bibinfo {author} {\bibfnamefont {M.}~\bibnamefont {Huang}},\ }\href
  {https://doi.org/10.1103/PhysRevD.108.096007} {\bibfield  {journal} {\bibinfo
   {journal} {Phys. Rev. D}\ }\textbf {\bibinfo {volume} {108}},\ \bibinfo
  {pages} {096007} (\bibinfo {year} {2023})},\ \Eprint
  {https://arxiv.org/abs/2307.14402} {arXiv:2307.14402 [hep-ph]} \BibitemShut
  {NoStop}%
\bibitem [{\citenamefont {Fujimoto}\ \emph {et~al.}(2021)\citenamefont
  {Fujimoto}, \citenamefont {Fukushima},\ and\ \citenamefont
  {Hidaka}}]{Fujimoto:2021xix}%
  \BibitemOpen
  \bibfield  {author} {\bibinfo {author} {\bibfnamefont {Y.}~\bibnamefont
  {Fujimoto}}, \bibinfo {author} {\bibfnamefont {K.}~\bibnamefont
  {Fukushima}},\ and\ \bibinfo {author} {\bibfnamefont {Y.}~\bibnamefont
  {Hidaka}},\ }\href {https://doi.org/10.1016/j.physletb.2021.136184}
  {\bibfield  {journal} {\bibinfo  {journal} {Phys. Lett. B}\ }\textbf
  {\bibinfo {volume} {816}},\ \bibinfo {pages} {136184} (\bibinfo {year}
  {2021})},\ \Eprint {https://arxiv.org/abs/2101.09173} {arXiv:2101.09173
  [hep-ph]} \BibitemShut {NoStop}%
\bibitem [{\citenamefont {Chen}\ \emph {et~al.}(2021)\citenamefont {Chen},
  \citenamefont {Zhang}, \citenamefont {Li}, \citenamefont {Hou},\ and\
  \citenamefont {Huang}}]{Chen:2020ath}%
  \BibitemOpen
  \bibfield  {author} {\bibinfo {author} {\bibfnamefont {X.}~\bibnamefont
  {Chen}}, \bibinfo {author} {\bibfnamefont {L.}~\bibnamefont {Zhang}},
  \bibinfo {author} {\bibfnamefont {D.}~\bibnamefont {Li}}, \bibinfo {author}
  {\bibfnamefont {D.}~\bibnamefont {Hou}},\ and\ \bibinfo {author}
  {\bibfnamefont {M.}~\bibnamefont {Huang}},\ }\href
  {https://doi.org/10.1007/JHEP07(2021)132} {\bibfield  {journal} {\bibinfo
  {journal} {JHEP}\ }\textbf {\bibinfo {volume} {07}},\ \bibinfo {pages}
  {132}},\ \Eprint {https://arxiv.org/abs/2010.14478} {arXiv:2010.14478
  [hep-ph]} \BibitemShut {NoStop}%
\bibitem [{\citenamefont {Braga}\ \emph {et~al.}(2022)\citenamefont {Braga},
  \citenamefont {Faulhaber},\ and\ \citenamefont {Junqueira}}]{Braga:2022yfe}%
  \BibitemOpen
  \bibfield  {author} {\bibinfo {author} {\bibfnamefont {N.~R.~F.}\
  \bibnamefont {Braga}}, \bibinfo {author} {\bibfnamefont {L.~F.}\ \bibnamefont
  {Faulhaber}},\ and\ \bibinfo {author} {\bibfnamefont {O.~C.}\ \bibnamefont
  {Junqueira}},\ }\href {https://doi.org/10.1103/PhysRevD.105.106003}
  {\bibfield  {journal} {\bibinfo  {journal} {Phys. Rev. D}\ }\textbf {\bibinfo
  {volume} {105}},\ \bibinfo {pages} {106003} (\bibinfo {year} {2022})},\
  \Eprint {https://arxiv.org/abs/2201.05581} {arXiv:2201.05581 [hep-th]}
  \BibitemShut {NoStop}%
\bibitem [{\citenamefont {Yadav}(2023)}]{Yadav:2022qcl}%
  \BibitemOpen
  \bibfield  {author} {\bibinfo {author} {\bibfnamefont {G.}~\bibnamefont
  {Yadav}},\ }\href {https://doi.org/10.1016/j.physletb.2023.137925} {\bibfield
   {journal} {\bibinfo  {journal} {Phys. Lett. B}\ }\textbf {\bibinfo {volume}
  {841}},\ \bibinfo {pages} {137925} (\bibinfo {year} {2023})},\ \Eprint
  {https://arxiv.org/abs/2203.11959} {arXiv:2203.11959 [hep-th]} \BibitemShut
  {NoStop}%
\bibitem [{\citenamefont {Wang}\ and\ \citenamefont
  {Feng}(2024)}]{Wang:2024szr}%
  \BibitemOpen
  \bibfield  {author} {\bibinfo {author} {\bibfnamefont {J.-H.}\ \bibnamefont
  {Wang}}\ and\ \bibinfo {author} {\bibfnamefont {S.-Q.}\ \bibnamefont
  {Feng}},\ }\href@noop {} {\  (\bibinfo {year} {2024})},\ \Eprint
  {https://arxiv.org/abs/2403.01814} {arXiv:2403.01814 [hep-ph]} \BibitemShut
  {NoStop}%
\bibitem [{\citenamefont {Yamamoto}\ and\ \citenamefont
  {Hirono}(2013)}]{Yamamoto:2013zwa}%
  \BibitemOpen
  \bibfield  {author} {\bibinfo {author} {\bibfnamefont {A.}~\bibnamefont
  {Yamamoto}}\ and\ \bibinfo {author} {\bibfnamefont {Y.}~\bibnamefont
  {Hirono}},\ }\href {https://doi.org/10.1103/PhysRevLett.111.081601}
  {\bibfield  {journal} {\bibinfo  {journal} {Phys. Rev. Lett.}\ }\textbf
  {\bibinfo {volume} {111}},\ \bibinfo {pages} {081601} (\bibinfo {year}
  {2013})},\ \Eprint {https://arxiv.org/abs/1303.6292} {arXiv:1303.6292
  [hep-lat]} \BibitemShut {NoStop}%
\bibitem [{\citenamefont {Braguta}\ \emph {et~al.}(2020)\citenamefont
  {Braguta}, \citenamefont {Kotov}, \citenamefont {Kuznedelev},\ and\
  \citenamefont {Roenko}}]{Braguta:2020biu}%
  \BibitemOpen
  \bibfield  {author} {\bibinfo {author} {\bibfnamefont {V.~V.}\ \bibnamefont
  {Braguta}}, \bibinfo {author} {\bibfnamefont {A.~Y.}\ \bibnamefont {Kotov}},
  \bibinfo {author} {\bibfnamefont {D.~D.}\ \bibnamefont {Kuznedelev}},\ and\
  \bibinfo {author} {\bibfnamefont {A.~A.}\ \bibnamefont {Roenko}},\ }\href
  {https://doi.org/10.31857/S1234567820130029} {\bibfield  {journal} {\bibinfo
  {journal} {Pisma Zh. Eksp. Teor. Fiz.}\ }\textbf {\bibinfo {volume} {112}},\
  \bibinfo {pages} {9} (\bibinfo {year} {2020})}\BibitemShut {NoStop}%
\bibitem [{\citenamefont {Braguta}\ \emph {et~al.}(2021)\citenamefont
  {Braguta}, \citenamefont {Kotov}, \citenamefont {Kuznedelev},\ and\
  \citenamefont {Roenko}}]{Braguta:2021jgn}%
  \BibitemOpen
  \bibfield  {author} {\bibinfo {author} {\bibfnamefont {V.~V.}\ \bibnamefont
  {Braguta}}, \bibinfo {author} {\bibfnamefont {A.~Y.}\ \bibnamefont {Kotov}},
  \bibinfo {author} {\bibfnamefont {D.~D.}\ \bibnamefont {Kuznedelev}},\ and\
  \bibinfo {author} {\bibfnamefont {A.~A.}\ \bibnamefont {Roenko}},\ }\href
  {https://doi.org/10.1103/PhysRevD.103.094515} {\bibfield  {journal} {\bibinfo
   {journal} {Phys. Rev. D}\ }\textbf {\bibinfo {volume} {103}},\ \bibinfo
  {pages} {094515} (\bibinfo {year} {2021})},\ \Eprint
  {https://arxiv.org/abs/2102.05084} {arXiv:2102.05084 [hep-lat]} \BibitemShut
  {NoStop}%
\bibitem [{\citenamefont {Yang}\ and\ \citenamefont
  {Huang}(2023)}]{Yang:2023vsw}%
  \BibitemOpen
  \bibfield  {author} {\bibinfo {author} {\bibfnamefont {J.-C.}\ \bibnamefont
  {Yang}}\ and\ \bibinfo {author} {\bibfnamefont {X.-G.}\ \bibnamefont
  {Huang}},\ }\href@noop {} {\  (\bibinfo {year} {2023})},\ \Eprint
  {https://arxiv.org/abs/2307.05755} {arXiv:2307.05755 [hep-lat]} \BibitemShut
  {NoStop}%
\bibitem [{\citenamefont {Mameda}\ and\ \citenamefont
  {Takizawa}(2023)}]{Mameda:2023sst}%
  \BibitemOpen
  \bibfield  {author} {\bibinfo {author} {\bibfnamefont {K.}~\bibnamefont
  {Mameda}}\ and\ \bibinfo {author} {\bibfnamefont {K.}~\bibnamefont
  {Takizawa}},\ }\href {https://doi.org/10.1016/j.physletb.2023.138317}
  {\bibfield  {journal} {\bibinfo  {journal} {Phys. Lett. B}\ }\textbf
  {\bibinfo {volume} {847}},\ \bibinfo {pages} {138317} (\bibinfo {year}
  {2023})},\ \Eprint {https://arxiv.org/abs/2308.07310} {arXiv:2308.07310
  [hep-ph]} \BibitemShut {NoStop}%
\bibitem [{\citenamefont {Sun}\ \emph {et~al.}(2024)\citenamefont {Sun},
  \citenamefont {Shao}, \citenamefont {Wen}, \citenamefont {Xu},\ and\
  \citenamefont {Huang}}]{Sun:2024anu}%
  \BibitemOpen
  \bibfield  {author} {\bibinfo {author} {\bibfnamefont {F.}~\bibnamefont
  {Sun}}, \bibinfo {author} {\bibfnamefont {J.}~\bibnamefont {Shao}}, \bibinfo
  {author} {\bibfnamefont {R.}~\bibnamefont {Wen}}, \bibinfo {author}
  {\bibfnamefont {K.}~\bibnamefont {Xu}},\ and\ \bibinfo {author}
  {\bibfnamefont {M.}~\bibnamefont {Huang}},\ }\href@noop {} {\  (\bibinfo
  {year} {2024})},\ \Eprint {https://arxiv.org/abs/2402.16595}
  {arXiv:2402.16595 [hep-ph]} \BibitemShut {NoStop}%
\bibitem [{\citenamefont {Cao}(2024)}]{Cao:2023olg}%
  \BibitemOpen
  \bibfield  {author} {\bibinfo {author} {\bibfnamefont {G.}~\bibnamefont
  {Cao}},\ }\href {https://doi.org/10.1103/PhysRevD.109.014001} {\bibfield
  {journal} {\bibinfo  {journal} {Phys. Rev. D}\ }\textbf {\bibinfo {volume}
  {109}},\ \bibinfo {pages} {014001} (\bibinfo {year} {2024})},\ \Eprint
  {https://arxiv.org/abs/2310.03310} {arXiv:2310.03310 [nucl-th]} \BibitemShut
  {NoStop}%
\bibitem [{\citenamefont {Jiang}(2023)}]{Jiang:2023hdr}%
  \BibitemOpen
  \bibfield  {author} {\bibinfo {author} {\bibfnamefont {Y.}~\bibnamefont
  {Jiang}},\ }\href@noop {} {\  (\bibinfo {year} {2023})},\ \Eprint
  {https://arxiv.org/abs/2312.06166} {arXiv:2312.06166 [hep-th]} \BibitemShut
  {NoStop}%
\bibitem [{\citenamefont {Gaspar}\ \emph {et~al.}(2023)\citenamefont {Gaspar},
  \citenamefont {Hern\'andez},\ and\ \citenamefont {Zamora}}]{Gaspar:2023nqk}%
  \BibitemOpen
  \bibfield  {author} {\bibinfo {author} {\bibfnamefont {I.~I.}\ \bibnamefont
  {Gaspar}}, \bibinfo {author} {\bibfnamefont {L.~A.}\ \bibnamefont
  {Hern\'andez}},\ and\ \bibinfo {author} {\bibfnamefont {R.}~\bibnamefont
  {Zamora}},\ }\href {https://doi.org/10.1103/PhysRevD.108.094020} {\bibfield
  {journal} {\bibinfo  {journal} {Phys. Rev. D}\ }\textbf {\bibinfo {volume}
  {108}},\ \bibinfo {pages} {094020} (\bibinfo {year} {2023})},\ \Eprint
  {https://arxiv.org/abs/2305.00101} {arXiv:2305.00101 [hep-ph]} \BibitemShut
  {NoStop}%
\bibitem [{\citenamefont {Tabatabaee~Mehr}(2023)}]{TabatabaeeMehr:2023tpt}%
  \BibitemOpen
  \bibfield  {author} {\bibinfo {author} {\bibfnamefont {S.~M.~A.}\
  \bibnamefont {Tabatabaee~Mehr}},\ }\href
  {https://doi.org/10.1103/PhysRevD.108.094042} {\bibfield  {journal} {\bibinfo
   {journal} {Phys. Rev. D}\ }\textbf {\bibinfo {volume} {108}},\ \bibinfo
  {pages} {094042} (\bibinfo {year} {2023})},\ \Eprint
  {https://arxiv.org/abs/2306.11753} {arXiv:2306.11753 [nucl-th]} \BibitemShut
  {NoStop}%
\bibitem [{\citenamefont {Chernodub}(2021)}]{Chernodub:2020qah}%
  \BibitemOpen
  \bibfield  {author} {\bibinfo {author} {\bibfnamefont {M.~N.}\ \bibnamefont
  {Chernodub}},\ }\href {https://doi.org/10.1103/PhysRevD.103.054027}
  {\bibfield  {journal} {\bibinfo  {journal} {Phys. Rev. D}\ }\textbf {\bibinfo
  {volume} {103}},\ \bibinfo {pages} {054027} (\bibinfo {year} {2021})},\
  \Eprint {https://arxiv.org/abs/2012.04924} {arXiv:2012.04924 [hep-ph]}
  \BibitemShut {NoStop}%
\bibitem [{\citenamefont {Chernodub}\ \emph {et~al.}(2023)\citenamefont
  {Chernodub}, \citenamefont {Goy},\ and\ \citenamefont
  {Molochkov}}]{Chernodub:2022veq}%
  \BibitemOpen
  \bibfield  {author} {\bibinfo {author} {\bibfnamefont {M.~N.}\ \bibnamefont
  {Chernodub}}, \bibinfo {author} {\bibfnamefont {V.~A.}\ \bibnamefont {Goy}},\
  and\ \bibinfo {author} {\bibfnamefont {A.~V.}\ \bibnamefont {Molochkov}},\
  }\href {https://doi.org/10.1103/PhysRevD.107.114502} {\bibfield  {journal}
  {\bibinfo  {journal} {Phys. Rev. D}\ }\textbf {\bibinfo {volume} {107}},\
  \bibinfo {pages} {114502} (\bibinfo {year} {2023})},\ \Eprint
  {https://arxiv.org/abs/2209.15534} {arXiv:2209.15534 [hep-lat]} \BibitemShut
  {NoStop}%
\bibitem [{\citenamefont {Braguta}\ \emph {et~al.}(2023)\citenamefont
  {Braguta}, \citenamefont {Chernodub},\ and\ \citenamefont
  {Roenko}}]{Braguta:2023iyx}%
  \BibitemOpen
  \bibfield  {author} {\bibinfo {author} {\bibfnamefont {V.~V.}\ \bibnamefont
  {Braguta}}, \bibinfo {author} {\bibfnamefont {M.~N.}\ \bibnamefont
  {Chernodub}},\ and\ \bibinfo {author} {\bibfnamefont {A.~A.}\ \bibnamefont
  {Roenko}},\ }\href@noop {} {\  (\bibinfo {year} {2023})},\ \Eprint
  {https://arxiv.org/abs/2312.13994} {arXiv:2312.13994 [hep-lat]} \BibitemShut
  {NoStop}%
\bibitem [{\citenamefont {Chen}\ \emph {et~al.}(2016)\citenamefont {Chen},
  \citenamefont {Fukushima}, \citenamefont {Huang},\ and\ \citenamefont
  {Mameda}}]{Chen:2015hfc}%
  \BibitemOpen
  \bibfield  {author} {\bibinfo {author} {\bibfnamefont {H.-L.}\ \bibnamefont
  {Chen}}, \bibinfo {author} {\bibfnamefont {K.}~\bibnamefont {Fukushima}},
  \bibinfo {author} {\bibfnamefont {X.-G.}\ \bibnamefont {Huang}},\ and\
  \bibinfo {author} {\bibfnamefont {K.}~\bibnamefont {Mameda}},\ }\href
  {https://doi.org/10.1103/PhysRevD.93.104052} {\bibfield  {journal} {\bibinfo
  {journal} {Phys. Rev. D}\ }\textbf {\bibinfo {volume} {93}},\ \bibinfo
  {pages} {104052} (\bibinfo {year} {2016})},\ \Eprint
  {https://arxiv.org/abs/1512.08974} {arXiv:1512.08974 [hep-ph]} \BibitemShut
  {NoStop}%
\bibitem [{\citenamefont {Flachi}\ and\ \citenamefont
  {Fukushima}(2018)}]{Flachi:2017vlp}%
  \BibitemOpen
  \bibfield  {author} {\bibinfo {author} {\bibfnamefont {A.}~\bibnamefont
  {Flachi}}\ and\ \bibinfo {author} {\bibfnamefont {K.}~\bibnamefont
  {Fukushima}},\ }\href {https://doi.org/10.1103/PhysRevD.98.096011} {\bibfield
   {journal} {\bibinfo  {journal} {Phys. Rev. D}\ }\textbf {\bibinfo {volume}
  {98}},\ \bibinfo {pages} {096011} (\bibinfo {year} {2018})},\ \Eprint
  {https://arxiv.org/abs/1702.04753} {arXiv:1702.04753 [hep-th]} \BibitemShut
  {NoStop}%
\bibitem [{\citenamefont {Zhang}\ \emph {et~al.}(2020)\citenamefont {Zhang},
  \citenamefont {Hou},\ and\ \citenamefont {Liao}}]{Zhang:2018ome}%
  \BibitemOpen
  \bibfield  {author} {\bibinfo {author} {\bibfnamefont {H.}~\bibnamefont
  {Zhang}}, \bibinfo {author} {\bibfnamefont {D.}~\bibnamefont {Hou}},\ and\
  \bibinfo {author} {\bibfnamefont {J.}~\bibnamefont {Liao}},\ }\href
  {https://doi.org/10.1088/1674-1137/abae4d} {\bibfield  {journal} {\bibinfo
  {journal} {Chin. Phys. C}\ }\textbf {\bibinfo {volume} {44}},\ \bibinfo
  {pages} {111001} (\bibinfo {year} {2020})},\ \Eprint
  {https://arxiv.org/abs/1812.11787} {arXiv:1812.11787 [hep-ph]} \BibitemShut
  {NoStop}%
\bibitem [{\citenamefont {Fukushima}\ and\ \citenamefont
  {Hatsuda}(2011)}]{Fukushima:2010bq}%
  \BibitemOpen
  \bibfield  {author} {\bibinfo {author} {\bibfnamefont {K.}~\bibnamefont
  {Fukushima}}\ and\ \bibinfo {author} {\bibfnamefont {T.}~\bibnamefont
  {Hatsuda}},\ }\href {https://doi.org/10.1088/0034-4885/74/1/014001}
  {\bibfield  {journal} {\bibinfo  {journal} {Rept. Prog. Phys.}\ }\textbf
  {\bibinfo {volume} {74}},\ \bibinfo {pages} {014001} (\bibinfo {year}
  {2011})},\ \Eprint {https://arxiv.org/abs/1005.4814} {arXiv:1005.4814
  [hep-ph]} \BibitemShut {NoStop}%
\bibitem [{\citenamefont {Adamczyk}\ \emph {et~al.}(2017)\citenamefont
  {Adamczyk} \emph {et~al.}}]{STAR:2017ckg}%
  \BibitemOpen
  \bibfield  {author} {\bibinfo {author} {\bibfnamefont {L.}~\bibnamefont
  {Adamczyk}} \emph {et~al.} (\bibinfo {collaboration} {STAR}),\ }\href
  {https://doi.org/10.1038/nature23004} {\bibfield  {journal} {\bibinfo
  {journal} {Nature}\ }\textbf {\bibinfo {volume} {548}},\ \bibinfo {pages}
  {62} (\bibinfo {year} {2017})},\ \Eprint {https://arxiv.org/abs/1701.06657}
  {arXiv:1701.06657 [nucl-ex]} \BibitemShut {NoStop}%
\bibitem [{\citenamefont {Fukushima}\ \emph {et~al.}(2020)\citenamefont
  {Fukushima}, \citenamefont {Shimazaki},\ and\ \citenamefont
  {Wang}}]{Fukushima:2020ncb}%
  \BibitemOpen
  \bibfield  {author} {\bibinfo {author} {\bibfnamefont {K.}~\bibnamefont
  {Fukushima}}, \bibinfo {author} {\bibfnamefont {T.}~\bibnamefont
  {Shimazaki}},\ and\ \bibinfo {author} {\bibfnamefont {L.}~\bibnamefont
  {Wang}},\ }\href {https://doi.org/10.1103/PhysRevD.102.014045} {\bibfield
  {journal} {\bibinfo  {journal} {Phys. Rev. D}\ }\textbf {\bibinfo {volume}
  {102}},\ \bibinfo {pages} {014045} (\bibinfo {year} {2020})},\ \Eprint
  {https://arxiv.org/abs/2004.05852} {arXiv:2004.05852 [hep-ph]} \BibitemShut
  {NoStop}%
\bibitem [{\citenamefont {Chen}\ \emph {et~al.}(2022)\citenamefont {Chen},
  \citenamefont {Fukushima},\ and\ \citenamefont {Shimada}}]{Chen:2022smf}%
  \BibitemOpen
  \bibfield  {author} {\bibinfo {author} {\bibfnamefont {S.}~\bibnamefont
  {Chen}}, \bibinfo {author} {\bibfnamefont {K.}~\bibnamefont {Fukushima}},\
  and\ \bibinfo {author} {\bibfnamefont {Y.}~\bibnamefont {Shimada}},\ }\href
  {https://doi.org/10.1103/PhysRevLett.129.242002} {\bibfield  {journal}
  {\bibinfo  {journal} {Phys. Rev. Lett.}\ }\textbf {\bibinfo {volume} {129}},\
  \bibinfo {pages} {242002} (\bibinfo {year} {2022})},\ \Eprint
  {https://arxiv.org/abs/2207.12665} {arXiv:2207.12665 [hep-ph]} \BibitemShut
  {NoStop}%
\bibitem [{\citenamefont {Gross}\ \emph {et~al.}(1981)\citenamefont {Gross},
  \citenamefont {Pisarski},\ and\ \citenamefont {Yaffe}}]{Gross:1980br}%
  \BibitemOpen
  \bibfield  {author} {\bibinfo {author} {\bibfnamefont {D.~J.}\ \bibnamefont
  {Gross}}, \bibinfo {author} {\bibfnamefont {R.~D.}\ \bibnamefont
  {Pisarski}},\ and\ \bibinfo {author} {\bibfnamefont {L.~G.}\ \bibnamefont
  {Yaffe}},\ }\href {https://doi.org/10.1103/RevModPhys.53.43} {\bibfield
  {journal} {\bibinfo  {journal} {Rev. Mod. Phys.}\ }\textbf {\bibinfo {volume}
  {53}},\ \bibinfo {pages} {43} (\bibinfo {year} {1981})}\BibitemShut {NoStop}%
\bibitem [{\citenamefont {Weiss}(1981)}]{Weiss:1980rj}%
  \BibitemOpen
  \bibfield  {author} {\bibinfo {author} {\bibfnamefont {N.}~\bibnamefont
  {Weiss}},\ }\href {https://doi.org/10.1103/PhysRevD.24.475} {\bibfield
  {journal} {\bibinfo  {journal} {Phys. Rev. D}\ }\textbf {\bibinfo {volume}
  {24}},\ \bibinfo {pages} {475} (\bibinfo {year} {1981})}\BibitemShut
  {NoStop}%
\bibitem [{\citenamefont {Weiss}(1982)}]{Weiss:1981ev}%
  \BibitemOpen
  \bibfield  {author} {\bibinfo {author} {\bibfnamefont {N.}~\bibnamefont
  {Weiss}},\ }\href {https://doi.org/10.1103/PhysRevD.25.2667} {\bibfield
  {journal} {\bibinfo  {journal} {Phys. Rev. D}\ }\textbf {\bibinfo {volume}
  {25}},\ \bibinfo {pages} {2667} (\bibinfo {year} {1982})}\BibitemShut
  {NoStop}%
\bibitem [{\citenamefont {Korthals~Altes}(1994)}]{KorthalsAltes:1993ca}%
  \BibitemOpen
  \bibfield  {author} {\bibinfo {author} {\bibfnamefont {C.~P.}\ \bibnamefont
  {Korthals~Altes}},\ }\href {https://doi.org/10.1016/0550-3213(94)90081-7}
  {\bibfield  {journal} {\bibinfo  {journal} {Nucl. Phys. B}\ }\textbf
  {\bibinfo {volume} {420}},\ \bibinfo {pages} {637} (\bibinfo {year}
  {1994})},\ \Eprint {https://arxiv.org/abs/hep-th/9310195}
  {arXiv:hep-th/9310195} \BibitemShut {NoStop}%
\bibitem [{\citenamefont {Gocksch}\ and\ \citenamefont
  {Pisarski}(1993)}]{Gocksch:1993iy}%
  \BibitemOpen
  \bibfield  {author} {\bibinfo {author} {\bibfnamefont {A.}~\bibnamefont
  {Gocksch}}\ and\ \bibinfo {author} {\bibfnamefont {R.~D.}\ \bibnamefont
  {Pisarski}},\ }\href {https://doi.org/10.1016/0550-3213(93)90123-7}
  {\bibfield  {journal} {\bibinfo  {journal} {Nucl. Phys. B}\ }\textbf
  {\bibinfo {volume} {402}},\ \bibinfo {pages} {657} (\bibinfo {year}
  {1993})},\ \Eprint {https://arxiv.org/abs/hep-ph/9302233}
  {arXiv:hep-ph/9302233} \BibitemShut {NoStop}%
\bibitem [{\citenamefont {Fukushima}\ and\ \citenamefont
  {Skokov}(2017)}]{Fukushima:2017csk}%
  \BibitemOpen
  \bibfield  {author} {\bibinfo {author} {\bibfnamefont {K.}~\bibnamefont
  {Fukushima}}\ and\ \bibinfo {author} {\bibfnamefont {V.}~\bibnamefont
  {Skokov}},\ }\href {https://doi.org/10.1016/j.ppnp.2017.05.002} {\bibfield
  {journal} {\bibinfo  {journal} {Prog. Part. Nucl. Phys.}\ }\textbf {\bibinfo
  {volume} {96}},\ \bibinfo {pages} {154} (\bibinfo {year} {2017})},\ \Eprint
  {https://arxiv.org/abs/1705.00718} {arXiv:1705.00718 [hep-ph]} \BibitemShut
  {NoStop}%
\bibitem [{\citenamefont {Chernodub}(2022)}]{Chernodub:2022qlz}%
  \BibitemOpen
  \bibfield  {author} {\bibinfo {author} {\bibfnamefont {M.~N.}\ \bibnamefont
  {Chernodub}},\ }\href@noop {} {\  (\bibinfo {year} {2022})},\ \Eprint
  {https://arxiv.org/abs/2210.05651} {arXiv:2210.05651 [quant-ph]} \BibitemShut
  {NoStop}%
\bibitem [{\citenamefont {Ebihara}\ \emph {et~al.}(2017)\citenamefont
  {Ebihara}, \citenamefont {Fukushima},\ and\ \citenamefont
  {Mameda}}]{Ebihara:2016fwa}%
  \BibitemOpen
  \bibfield  {author} {\bibinfo {author} {\bibfnamefont {S.}~\bibnamefont
  {Ebihara}}, \bibinfo {author} {\bibfnamefont {K.}~\bibnamefont {Fukushima}},\
  and\ \bibinfo {author} {\bibfnamefont {K.}~\bibnamefont {Mameda}},\ }\href
  {https://doi.org/10.1016/j.physletb.2016.11.010} {\bibfield  {journal}
  {\bibinfo  {journal} {Phys. Lett. B}\ }\textbf {\bibinfo {volume} {764}},\
  \bibinfo {pages} {94} (\bibinfo {year} {2017})},\ \Eprint
  {https://arxiv.org/abs/1608.00336} {arXiv:1608.00336 [hep-ph]} \BibitemShut
  {NoStop}%
\end{thebibliography}%
\bibliographystyle{apsrev4-2}
\end{document}